# Tuning the electronic and chemisorption properties of hexagonal MgO nanotubes by doping – Theoretical study

Aleksandar Jovanović[1,2], Milena Petković[1], Igor A. Pašti[1,3]*, Börje Johansson[3,4,5], Natalia V. Skorodumova[3,4]

[1]*University of Belgrade – Faculty of Physical Chemistry, Studentski trg 12-16, 11158 Belgrade, Serbia*

[2]*CEST Kompetenzzentrum für elektrochemische Oberflächentechnologie GmbH, Viktor-Kaplan-strasse 2, Section A, 2700 Wiener Neustadt, Austria*

[3]*Department of Materials Science and Engineering, School of Industrial Engineering and Management, KTH - Royal Institute of Technology, Brinellvägen 23, 100 44 Stockholm, Sweden*

[4]*Department of Physics and Astronomy, Uppsala University, Box 516, 751 20 Uppsala, Sweden*

[5]*Humboldt University, Physics Department, Zum Grossen Windkanal 6, D-12489 Berlin, Germany*

* **corresponding author:** e-mail: igor@ffh.bg.ac.rs

**Abstract**

Oxide materials offer a wide range of interesting physical and chemical properties. Even more versatile behavior of oxides is seen at the nanoscale, qualifying these materials for a number of applications. In this study we used DFT calculations to investigate the physical and chemical properties of small hexagonal MgO nanotubes of different length. We analyzed the effect of Li, B, C, N, and F doping on the properties of the nanotubes. We find that all dopants favor the edge positions when incorporated into the nanotubes. Doping results in the net magnetization whose value depends on the type of the impurity. Using the CO molecule as a probe, we studied the adsorption properties of pristine and doped MgO nanotubes. Our results show that the dopant sites are also the centers of significantly altered chemical reactivity. While pristine MgO nanotubes adsorb CO weakly, very strong adsorption at the dopant sites (B-, C-, and N-doped nanotubes) or neighboring edge atoms (F- and Li-doped nanotubes) is observed. Our results suggest that impurity engineering in oxide materials can be a promising strategy for the development of novel materials with possible use as selective adsorbents or catalysts.

## 1. Introduction

Oxides are formed by virtually all the elements of the Periodic Table. They provide a wide variety of properties, ensuring their application in many different fields [1,2]. The surface properties of oxides, which are of paramount importance when the reactivity is considered, depend on morphology, specific surface area, exposure of crystallographic planes, impurities and defects, and other factors [3,4]. Oxides can be used in their native forms, but they can also be functionalized, tuning their properties for a particular application [5–9].

Magnesium oxide which appears in the rock salt structure [10] is very stable, both thermally and chemically, and as such is applied in many areas, ranging from refractory materials [11] to catalysis [5,6]. Its most stable surface, MgO(001), shows very poor reactivity [12], but its properties can be modified using various strategies [13–18]. Recently, the $d^0$ magnetism has been observed in N-doped MgO films with N taking the position of O in the MgO lattice [19]. The appearance of magnetic moment has also been predicted by Shein *et al*. in the case of C-doped multi-walled MgO nanotubes, where the C atom is introduced into the lattice instead of oxygen [20]. The same authors suggested that the magnetic moment will appear with dopants having the valence bands above the O 2*p* band [20]. The results of a theoretical study done for B-, C-, and N- doped MgO(001) surfaces support these predictions [21]. At the same time, the mentioned work demonstrates that the dopant sites exhibit drastically enhanced reactivity, making such systems interesting for heterogenous catalysis or as adsorbents [21,22].

At the nanoscale, the properties of many materials, including oxides, differ drastically from those of the bulk materials. In particular, the variation of structural parameters in nanosized clusters can have an impact on their electronic structure, which, in turn, affects the reactivity of such systems. These factors must be considered in materials modeling, since some desired properties may only emerge in nanostructured systems. MgO clusters have intensively been studied, both experimentally and theoretically. Mass spectroscopy measurements performed on $(MgO)_{3k}$ clusters confirm the existence of clusters with $k$ between 1 and 10 [23,24].

The theoretical investigation of small $(MgO)_{3k}$ clusters shows that for $k < 6$, the hexagonal nanotubes are preferred over the rock salt structures [25]. Changes of the chemical properties of such MgO nanotubes ($k = 4$) was reported by Yang *et al*. [26] who showed that substitutional doping by Ni, Pd and Pt enhanced the CO adsorption properties of hexagonal $(MgO)_{12}$ ($k$ = 4) nanotubes. The mentioned dopants replaced Mg atoms in the MgO nanotubes that significantly affected the CO adsorption energies. Moreover, the reactivity of dopant sites was shown to depend on the type of dopant and the Mg site at which dopant was introduced into the $(MgO)_{12}$ nanotube, as less coordinated dopants were found to be more reactive [26].

In this study, we extend our previous work on non-metal doped MgO(001) [21] for the case of hexagonal MgO nanotubes and investigate how doping with Li, B, C, N, and F can tune their properties. Due to their atomic sizes these dopants are considered as suitable ones for MgO, bulk, surface or nanostructures, in particular, nanotubes addressed here. Namely, these atoms are close to Mg and O in the Periodic Table of Elements and can effectively replace O or Mg atoms in the MgO lattice. Moreover, N-doped MgO films have been prepared experimentally [19]. An additional motivation for the present research is very the very promising reactivity of doped MgO(001) surfaces, predicted theoretically [21,22]. In the present work we analyze the electronic and magnetic properties of these clusters and investigate their chemical reactivity using CO as a probe for molecular adsorption.

## 2. Computational details

The calculations were based on Density Functional Theory (DFT) using the Generalized Gradient Approximation in the parametrization of Perdew, Burke and Ernzerhof (PBE) [27]. The calculations were performed with the Quantum ESPRESSO (QE) *ab initio* package [28] using ultrasoft pseudopotentials. The *s*- and *p*-states of all the atoms were included into the valence band. The kinetic energy cutoff for the plane-wave basis set was 380 eV and the charge density cutoff was 16 times higher. Spin polarization was taken into account for all the investigated systems. The contribution of the long range dispersion interactions were investigated using the DFT+D2 scheme by Grimme [29]. In order to avoid the interaction of clusters with their periodic images, large supercells (distance between periodic images at least 10 Å) were used in combination with the Martyna-Tuckerman correction [30]. Also, the torque of the internal forces was set to zero to compensate for the interaction with the periodic images [28]. The Brillouin zone was sampled using the $\Gamma$-point. Relaxation was performed until the residual forces were below 0.01 eV Å$^{-1}$. The charge transfer was analyzed using the Bader algorithm [31] on a charge density grid by Henkelman *et al*. [32].

In addition to the calculations performed in QE, for selected systems we also performed a set of calculations employing hybrid functionals, using Gaussian 09 program package [33]. This allowed us to compare PBE and PBE+D2 results with those obtained using B3LYP functional [34,35] combined with the cc-pVTZ basis set [36,37]. The structures were optimized using the default convergence criteria, and the absence of imaginary frequencies confirmed that the true minima of the potential energy surfaces were identified.

The nanotubes were constructed by vertical stacking of 2, 3, 4, or 5 hexagonal $(MgO)_3$ rings. Fig. 1 shows the doped $(MgO)_{12}$ nanotube consisting of four $(MgO)_3$ rings. One dopant atom per nanotube was introduced, substituting an O atom in the case of B-, C-, N-, and F-doping, and a Mg atom in the case of Li-doping.

To address the energetics of the substitution of atoms in the MgO lattice with ~~a~~ dopant X (X = B, C, N, F, or Li) we define substitution energy, $E_{sub}(X)$, as:

$$E_{sub}(X) = (E_{X\text{-}NT} + E_{O/Mg}) - (E_{NT} + E_{X}) \tag{1}$$

In the equation above, $E_{X\text{-}NT}$ and $E_{NT}$ stand for the total energies of the X-doped and pristine nanotube, respectively. $E_{O/Mg}$ denotes the total energy of an isolated O or Mg atom, depending on the dopant, and $E_X$ is the total energy of an isolated dopant atom. Alternatively, the incorporation of X into the nanotube can be considered as the binding of atom X to a vacancy site, which can be quantified as binding energy, $E_b(X)$:

$$E_b(X) = E_{X\text{-}NT} - (E_{v\text{-}NT} + E_{X}) \tag{2}$$

where $E_{v\text{-}NT}$ denotes the total energy of the nanotube with a vacancy (O or Mg vacancy). Vacancy formation energy, $E_f^{vac}$, can be defined as:

$$E_f^{vac} = E_{v\text{-}NT} - (E_{NT} + E_{O/Mg}) \tag{3}$$

The adsorption properties of pristine and the X-doped nanotubes were probed using CO. The adsorption is quantified by the CO adsorption energy, $E_{ads}(CO)$, defined as:

$$E_{ads}(CO) = E_{CO@X\text{-}NT} - (E_{NT} + E_{CO}) \tag{4}$$

where $E_{CO@X\text{-}NT}$ stands for the total energy of the nanotube with the CO molecule and $E_{CO}$ is the total energy of an isolated CO molecule. For the doped nanotubes, we studied the CO adsorption only for the most stable configurations of the doped $(MgO)_{12}$ nanotubes. The notations for the adsorption sites are presented in Fig. 1.

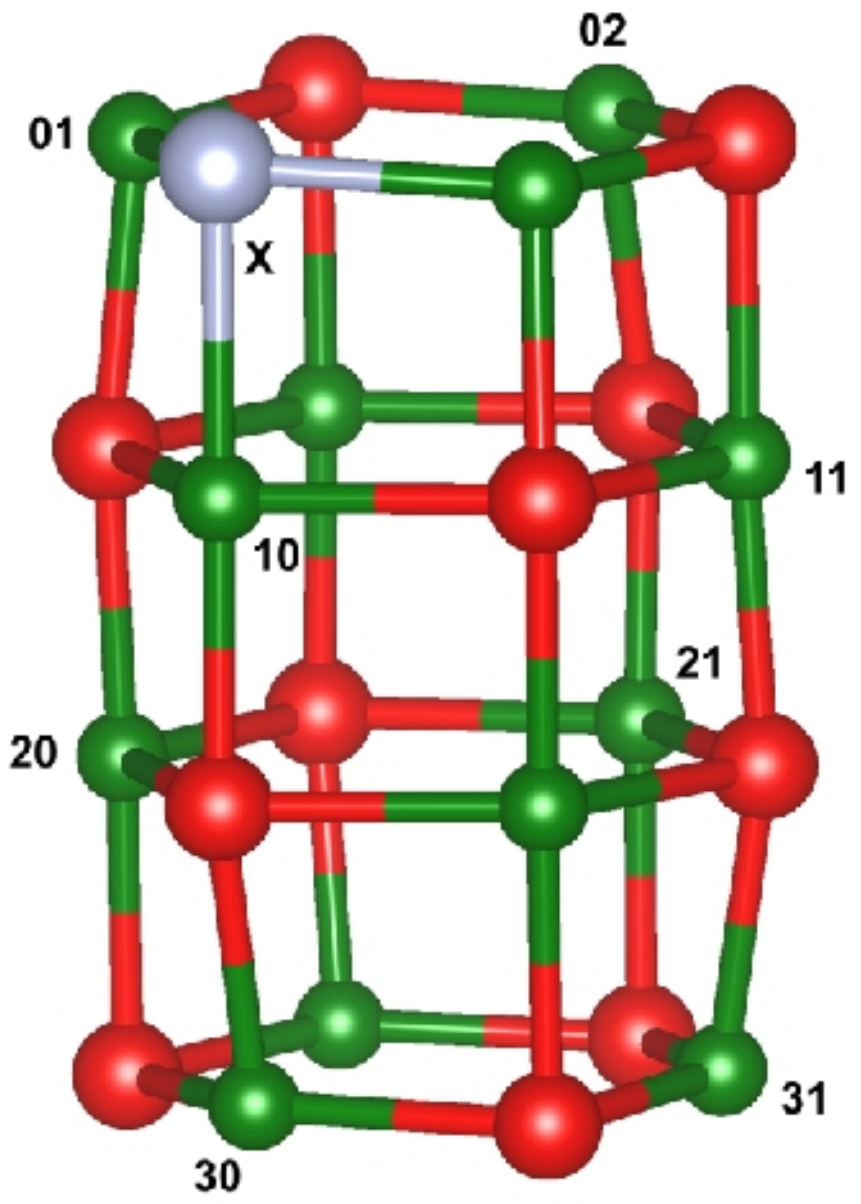

**Figure 1**. Notation of the adsorption sites at the doped MgO nanotube. Graphical presentation was made using the VMD code [44]. X denotes the dopant site.

## 3. Results

### 3.1. Structural, electronic, and magnetic properties of doped MgO nanotubes

The analysis of the hexagonal MgO nanotubes was performed for the nanotubes ranging from 2 to 5 rings in size. The Mg-O bonds in the nanotubes ranging between 1.90 and 2.05 Å (Supplementary Information, Tables S1 to S4) are somewhat shorter than that in the bulk (2.11 Å). Moreover, due to a lower coordination number of the edge sites, the reduction of the Mg-O bond is more noticeable at the nanotubes edges, in agreement with Ref. [25]. The bond lengths obtained in PBE and PBE+D2 calculations compare well with those calculated using B3LYP with the localized basis set. The comparison of the results obtained by these approaches for pristine MgO nanotubes is provided in Supplementary Information, together with the optimized structures (see Section S1).

The electronic structure of the investigated pristine MgO nanotubes was analyzed using both the periodic PBE calculations performed with QE and hybrid B3LYP calculations using Gaussian code. Fig. 2 shows the calculated densities of states (PBE calculations) and the orbital energy levels obtained using the B3LYP functional. As can be seen, the HOMO-LUMO gaps estimated from the results of the hybrid calculations are larger than the corresponding band gaps obtained in the periodic PBE calculations. Also, as shown in Fig. 2, the gap increases with the size of the MgO nanotubes.

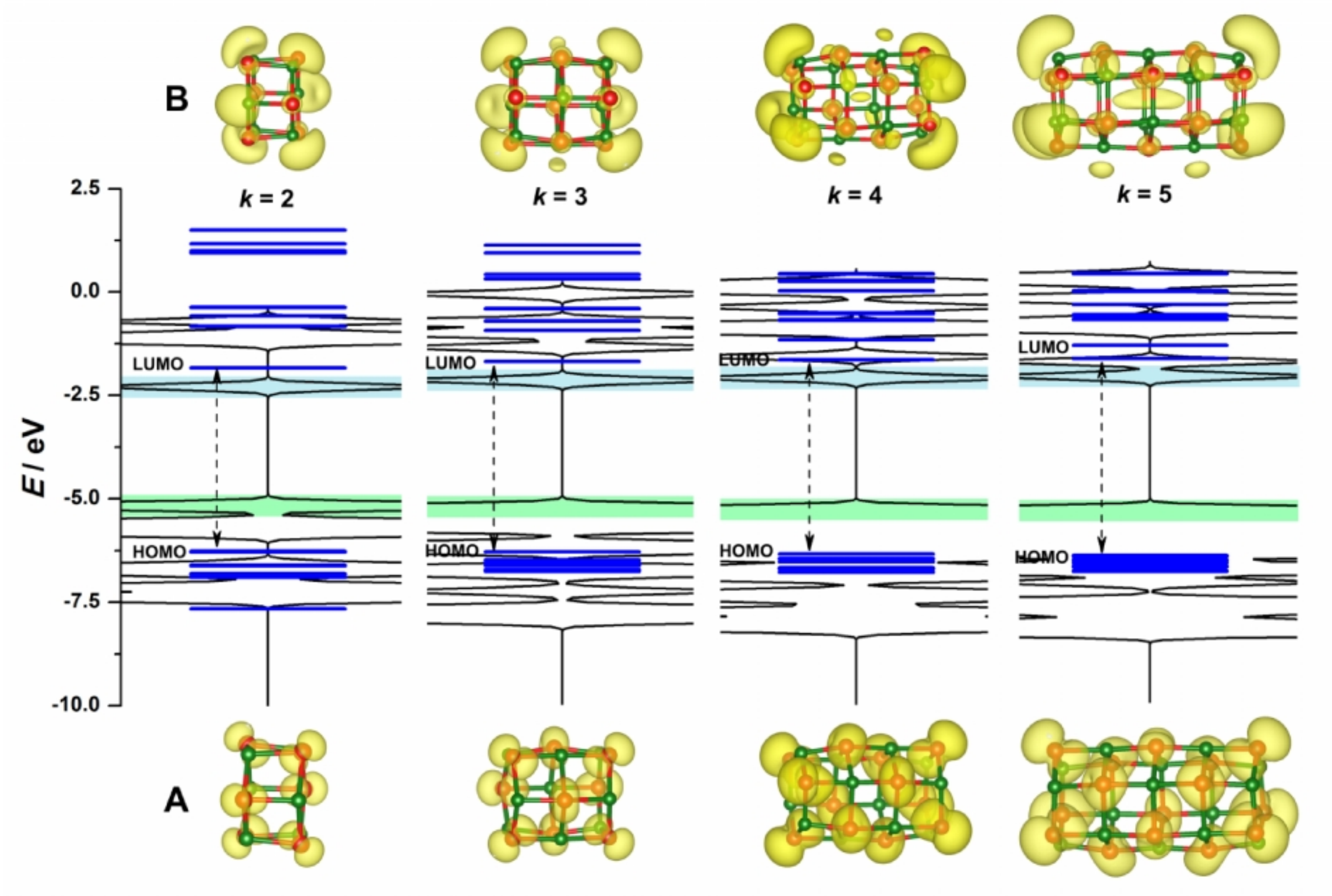

**Figure 2**. Calculated density of states using periodic PBE calculations in QE and the corresponding energy levels calculated using B3LYP with Gaussian code (blue horizontal lines). Vertical dashed arrows indicate HOMO-LUMO gap for the B3LYP calculations. The integrated local density of states (ILDOS) within the shaded areas is presented and it gives the top of valence (green shaded areas and the A set of the structures) and the bottom of conduction band (blue shaded areas and the B set of the structures) obtained using periodic PBE calculations in QE. The structures are presented using VESTA [45].

The HOMO-LUMO gaps calculated using B3LYP were found to increase from 4.44 eV, to 4.72 eV as the nanotube size increases from $(MgO)_6$ to $(MgO)_{15}$. The corresponding values obtained from the PBE calculations are between 2.48 eV, and 2.85 eV. These gaps are still much narrower than the one of bulk MgO, which amounts to 6.3 eV (for surface) and 7.8 eV (for bulk) [38]. For the $(MgO)_{12}$ nanotubes the band gap of 3.39 eV was previously reported [26], which is in-between our PBE and B3LYP values. We analyzed the electronic states of pristine hexagonal MgO nanotubes by visualizing the integrated local densities of states (ILDOS) of the top of the valence band and the bottom of the conduction band obtained in the periodic PBE calculations (Fig. 2). The former states are of oxygen atoms (Fig. 2, structures A), while the latter ones come from Mg atoms, predominantly those at the edges of the nanotubes (Fig. 2, structures B). This agrees with the results of hybrid calculations where we observed the HOMO states located at oxygen atoms, while the LUMO states were located at Mg atoms. We conclude that PBE and B3LYP give rather consistent results.

Before proceeding with the analysis of the doping process of $(MgO)_{3k}$ nanotubes, we discuss the effects of the edge saturation by hydrogen. Hydrogen saturation of dangling bonds is an important issue [39], however, in the previous studies of MgO nanotubular structures it has not been considered [20,24,25]. Also, there is no experimental evidence supporting the edge saturation [22,23]. In order to investigate this issue we modelled a number of H-saturated $(MgO)_{3k}$ clusters (Supplementary Information, Section S2). We observed that upon saturation of the edges of $(MgO)_{3k}$ clusters the edge Mg-O bonds become longer. However, the energetics of this process is much more important. Namely, when referred to $H_2$ or $H_2O$, as possible sources of H atoms, which saturate the edges of $(MgO)_{3k}$ clusters, we see that the saturation is a highly endothermic process (Supplementary Information, Section S2). What is why we do not consider the saturation of the edges of $(MgO)_{3k}$ clusters from this point on.

In order to incorporate an impurity atom into an MgO nanotube, first a vacancy has to be formed. We analyzed the energetics of the vacancy formation and found that it depended on the location of the vacancy in the nanotube structure. Due to the reduced coordination of atoms, it is always easier to make a vacancy at the edge of the nanotube. Moreover, we find that $E_f^{vac}$ increases with the size of the nanotube and when moving away from its edge (Fig. 3, the results for the formation of oxygen vacancy). This conclusion holds for both oxygen and magnesium vacancies (Table S5, Supplementary Information) and for PBE and PBE+D2 calculations. In the case of the oxygen vacancy we find somewhat higher contribution of dispersion interactions to the vacancy formation energy away from the nanotube edge, compared to the energies obtained for the vacancies at nanotube edges. However, the contribution of dispersion interactions is generally quite small (within 0.1 eV) when compared with the vacancy formation energies themselves (8.5 eV on average). In the case of the Mg vacancy, the required energy is somewhat lower (5.9 eV on average), but the contribution of dispersion is higher (0.2 to 0.35 eV) compared to the case of the oxygen vacancy formation. The agreement between the PBE(+D2) and B3LYP results is better in the case of the O vacancy than in the case of Mg vacancy (Table S5). In contrast the PBE results, in both cases B3LYP results show no trend in vacancy formation energies with increasing nanotube size (compare Fig. 5 and the results in Table S5).

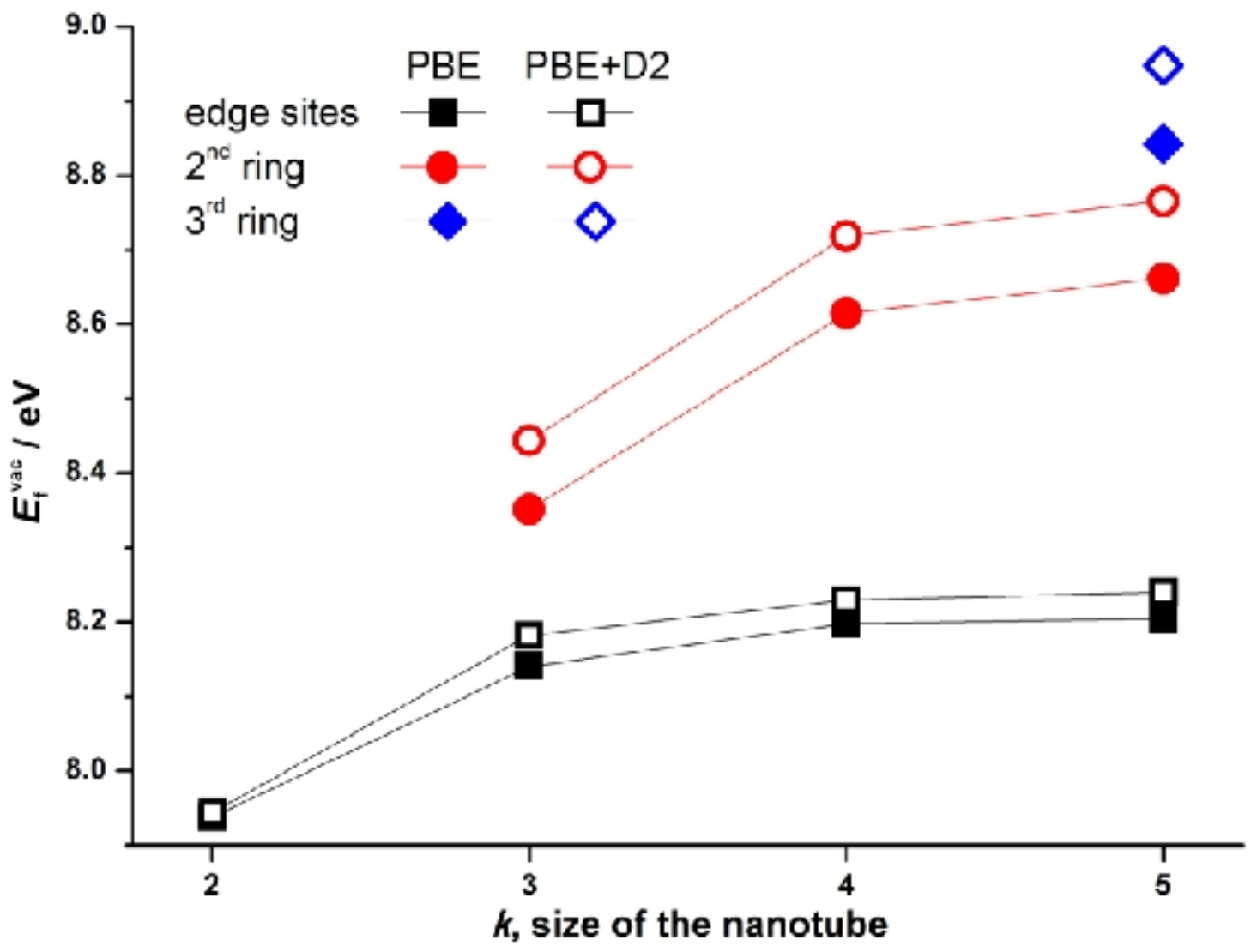

**Figure 3**. Oxygen vacancy formation energies for the MgO nanotubes of different length, defined by the number of $(MgO)_3$ rings ($k$). Squares give O vacancy formation energy at the edge, circles for the second $(MgO)_3$ ring, while the diamonds are for the third $(MgO)_3$ ring.

Next, we incorporated a dopant into the vacancy site. We replaced O with B, C, N, or F, or Mg with Li. The atoms were replaced at different sites along the nanotube, one atom *per* $(MgO)_{3k}$ nanotube. For all the considered dopants and the nanotube sizes (from $(MgO)_6$ to $(MgO)_{15}$) we found that dopants preferred the edge sites of the nanotube. This is in agreement with the results reported for the C-doped multiwall MgO nanotubes [20], and the MgO(001) terraces doped with B, C and N [21].

Moreover, we performed additional calculations for F- and Li doped Mg(001), using the model presented in Ref. [21], and confirmed that these two dopants prefer to be in the surface rather than subsurface layer. In order to evaluate the energetics of the doping process, we can either consider the replacement of a lattice atom with dopant X or binding of dopant X to the vacancy site. Dopant binding energies (Eq. 2) typically become more negative when going away from the nanotube edge, similarly to the increase of the vacancy formation energies for the higher coordinated sites. However, the most stable structure is determined by the balance between the vacancy formation energy and the binding energy of dopant X at a given vacancy site. As a result, we see that the configurations of doped MgO nanotubes containing dopants at the edge sites are most stable. Such a behavior can be understood considering the strain induced upon doping. Namely, due to the lower coordination of the edge atoms, strain is best compensated when the dopant is situated at the edge. The data for the $(MgO)_{12}$ nanotubes are provided in Table 1. The corresponding optimized structures (QE-PBE results) with indicated bond lengths are presented in Figure S1 (Supplementary Information). The calculated binding energies for all the considered dopants and nanotubes are assembled in Table S6. For dopants bound to the O vacancy site, the adsorption is more exothermic as the impurity approaches O in the Periodic Table. We also see that the substitution energies are positive in all the considered cases, which is the consequence of very high vacancy formation energies, when compared to the binding energies of dopants.

**Table 1**. Substitution energies, binding energies, and magnetic properties of B-, C-, N-, F, and Li- doped $(MgO)_{12}$ nanotubes. Binding energies are calculated as the binding of the atom X at the corresponding vacancy site.

| | $E_{sub}$(X) / eV | | | $E_b$(X) / eV | | | |
|---|---|---|---|---|---|---|---|
| **Dopant** | **PBE** | **PBE+D2** | **B3LYP** | **PBE** | **PBE+D2** | **B3LYP** | $M$ / $\mu_B$ |
| B | 5.08 | 5.09 | 5.31 | −3.11 | −3.15 | −3.36 | 3 |
| C | 3.67 | 3.66 | 3.64 | −4.54 | −4.56 | −5.03 | 2 |
| N | 3.65 | 3.65 | 3.19 | −4.55 | −4.58 | −5.48 | 1 |
| F | 2.05 | 2.02 | 1.69 | −6.20 | −6.21 | −6.99 | 1 |
| Li | 1.22 | 1.16 | 2.32 | −4.48 | −4.76 | −7.30 | 1 |

The contribution of dispersion interactions is almost negligible for both binding and substitution energies. We see rather good agreement (Table 1) between the PBE(+D2) results and the B3LYP ones. Although there are some differences in the calculated energies, they are in line with the previously reported trends for adsorption energetics on oxide surface when comparing GGA and hybrid functionals [40]. The difference between our PBE and B3LYP results is the largest in the case of Li-doping and it originates from the differences in the calculated Mg vacancy formation energies using these two approaches (Table S5).

For B, C, and N doping, the Bader charge analysis shows a charge transfer to the dopant, as previously also observed for the (001) surface doped with these elements [21]. This results in a significantly increased ionic radii and the increase of the Mg-X bond length, Fig. S2. It is the most apparent for B, and the effect decreases towards N, as discussed in Ref. [21]. For F and Li, having $2s^2np^5$ and $2s^1$ electron configuration, respectively, only one electron is transferred upon their incorporation into the MgO nanotube. F attracts extra charge, while Li gives its valence electron away, so the dopants are formally in the $F^-$ and $Li^+$ states. This, however, means that the surrounding lattice atoms (Mg or O) do not have electronic configurations as in pristine MgO nanotubes. Hence, in the case of Li-doped MgO nanotubes the O atoms around the dopant atom are electron deficient. In contrast, in the case of the F-doped MgO nanotubes neighboring Mg atoms possess an excess charge. Charge density plots (Supplementary Information, Fig. S3) show that the electronic density at the O sites is not significantly affected by the inclusion of dopant atoms. This suggests that the effect of the dopant atoms is rather localized around the impurity sites. In order to demonstrate this more clearly, we calculated the charge different plots upon the replacement of the edge O atom (in the cases of B, C, N and F) or Mg atom (in the case of Li) by a dopant in $(MgO)_{12}$ clusters. Charge difference plots are obtained as:

$$\Delta\rho = \rho_{X\text{-}MgO} - (\rho_{MgO} - \rho_{O/Mg}) - \rho_X \qquad (5)$$

In Eq. (5) $\rho_{X\text{-}MgO}$, $\rho_{MgO}$, $\rho_{O/Mg}$ and $\rho_X$ stand for the ground state charge density of the doped MgO nanotube, the ground state charge density of pristine MgO nanotube, the ground state charge density of O or Mg atom and the ground state charge density of dopant X, respectively. The results presented in Fig. 4 reinforce the conclusion that the replacement of an O or Mg atom by dopant X induces only the local charge redistribution in the MgO nanotube.

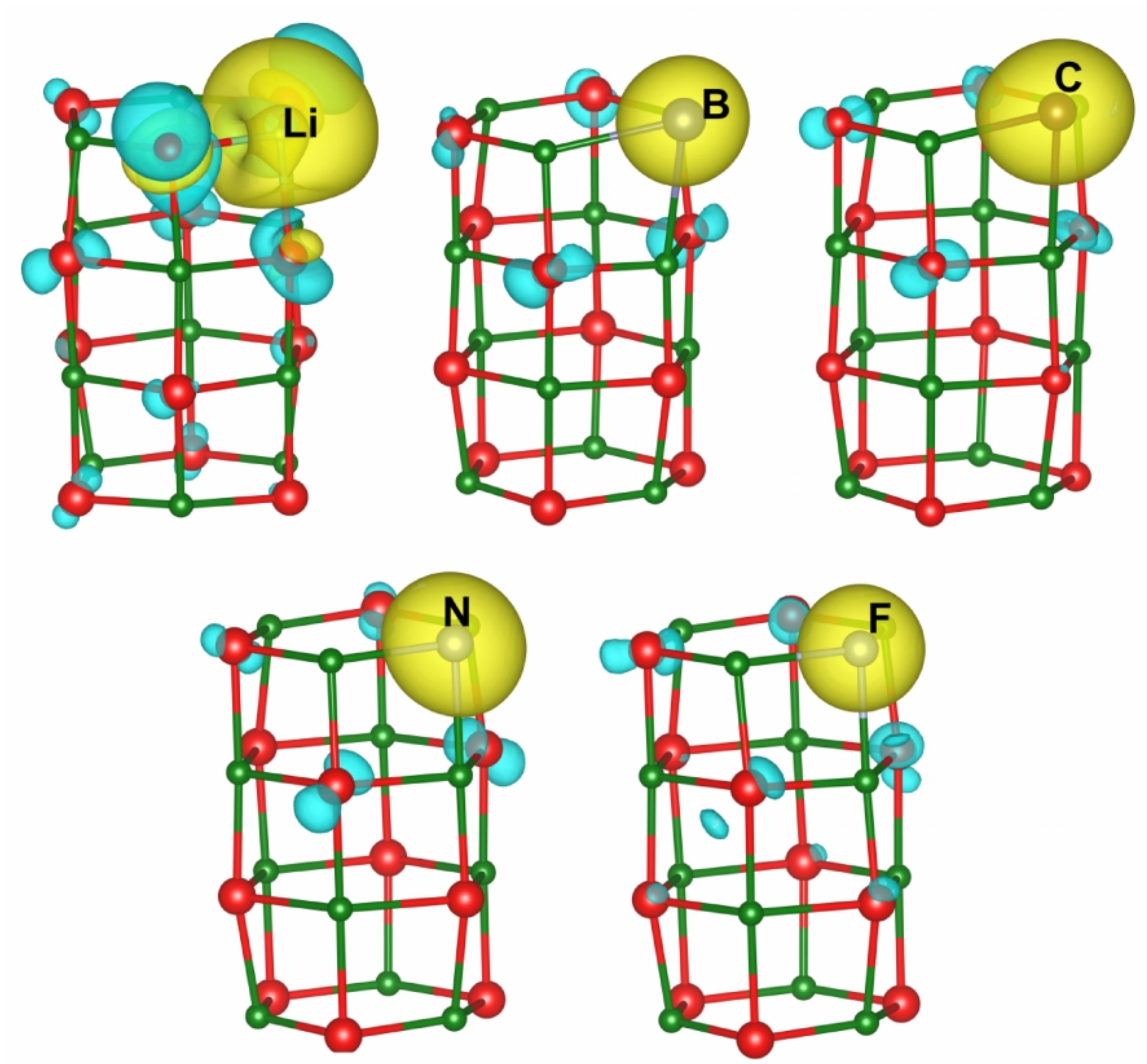

**Figure 4**. 3D charge density difference plots for replacement of the edge O or Mg atom in the $(MgO)_{12}$ nanotube by Li, B, C, N or F.

The charge transfer is also responsible for the appearance of a magnetic moment. The total magnetization values observed for B-, C-, and N-doped nanotubes are in agreement with those obtained for doped MgO(001) [21]. The magnetic moments originate from the unpaired electrons of dopant atoms independent of the nanotube size, but they rather depend on the type of the dopant atom (Fig. 5). Boron, with one unpaired electron, receives 2 extra electrons from the lattice and the total number of unpaired electrons becomes 3. It matches the total magnetization of 3 $\mu_B$ we obtained for the B-doped MgO nanotubes. Similar behavior is observed for C and N doping where the total magnetization is 2 and 1 $\mu_B$, respectively. The calculated total magnetizations agree well with those provided by Shein *et al*. for the C-doped triple-walled square-prismatic MgO nanotubes [20], as well as with the calculations of Grob *et al*. for N-doped MgO [19]. Since F can only receive 1 electron to completely fill its *p*-shell, the magnetization is now due to an excess charge on the surrounding Mg atoms, amounting to 1 $\mu_B$ (Fig. 5). The situation is similar for Li doping, where the magnetization comes from the electron-deficient O atoms surrounding the dopant (Fig. 5). It can be concluded that the magnetization density in the investigated systems is rather localized around the impurity site (Fig. 5).

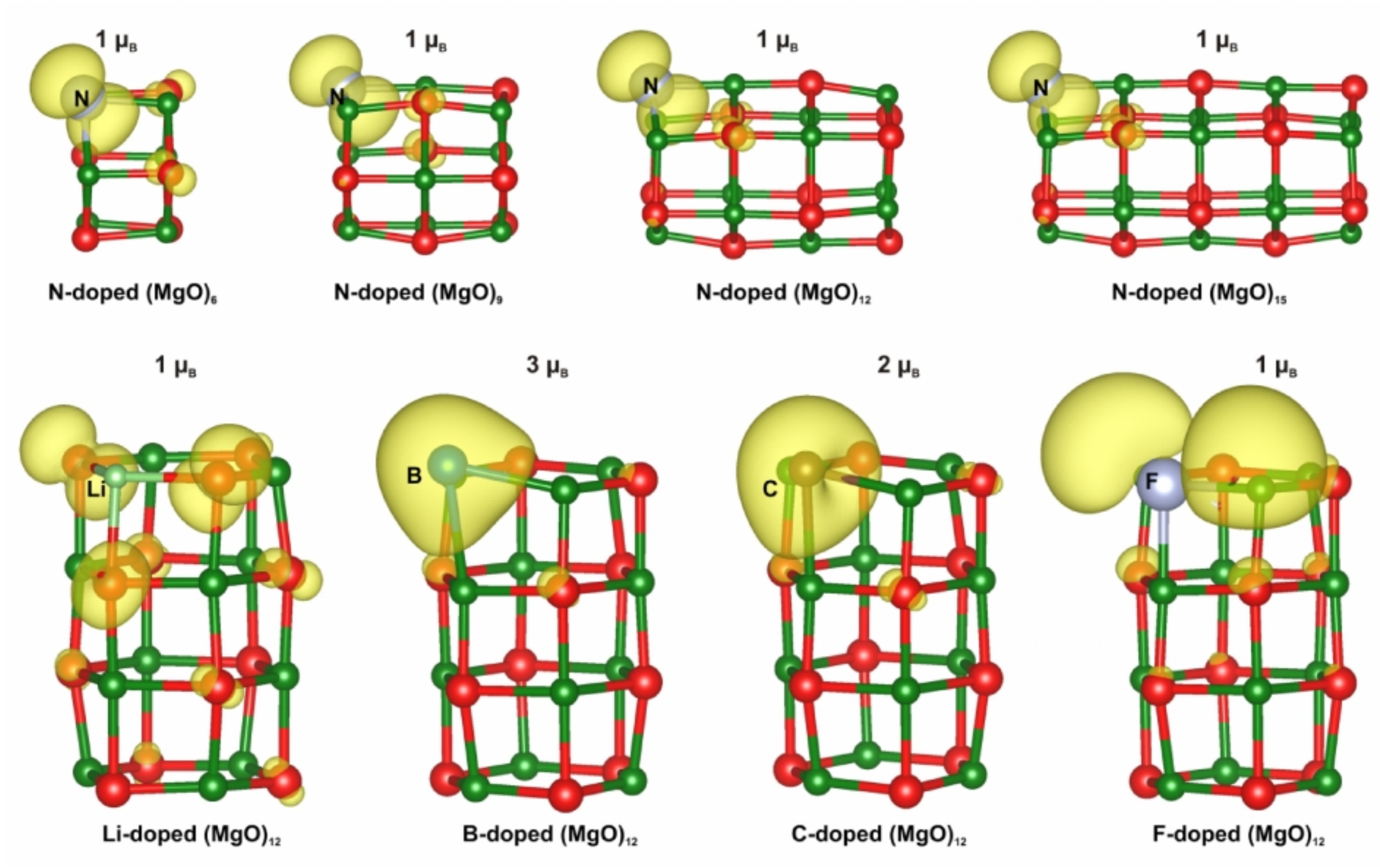

**Figure 5**. 3D spin density (spin polarization) maps ($\rho_{spin\ up}$ - $\rho_{spin\ down}$). Calculated total magnetizations are given for each structure.

The electronic structures of the doped $(MgO)_{12}$ nanotubes, shown in Fig. 6, demonstrate that the magnetization is due to the charge redistribution between the dopant atoms and lattice atoms of the MgO nanotube. For B, C, and N dopants the calculated electronic structures of the doped nanotubes qualitatively agree with the results for the doped bulk and (001) surface of MgO [19,21]. The spin up *p* states of the dopant atoms are filled, while the spin down states are partially empty, giving rise to the net magnetic moment, in agreement with Ref. [20]. In the case of the Li-doped MgO nanotubes the valence 2*s* states of the dopant are emptied while the top of the valence band is due to the stated of oxygen atoms (Fig. 6). The magnetic moment is due to the hole states located at the O atoms in the first coordination shell of Li (Fig. 5). In the case of the F-doped nanotubes, the electrons, which are located at the nearest Mg sites, give rise to the magnetic moment (Fig. 5). These states are located at the bottom of the conduction band of the F-doped nanotube (Fig. 6). The states induced by dopants also alter the band gap of MgO nanotubes. These states are typically located in the band gap, like in the case of doped MgO(001) surface [21] that results in gap narrowing. In fact, Li and F doping present two extreme cases. In the first case the band gap is 2.58 eV, being just slightly affected by doping while in the case of F doping the gap is closed (Fig. 6). In the cases of B, C and N doping the dopant states are between the valence and conduction bands and HOMO-LUMO gaps can be estimated to 0.68 eV (B doping), 0.84 eV (C doping) and 0.90 eV (N doping).

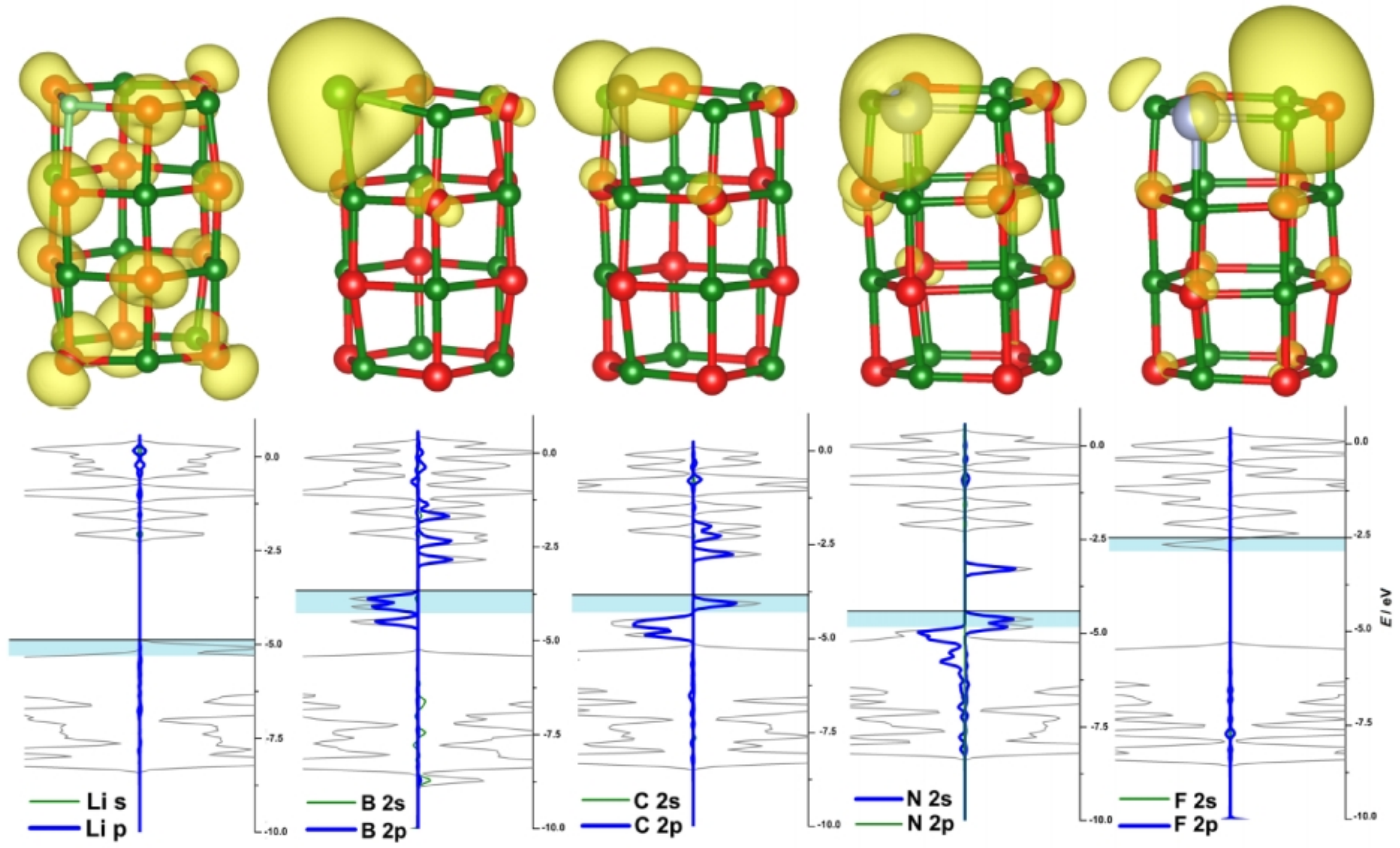

**Figure 6**. Densities of States of X-doped nanotubes (total DOS, and the *s* and *p* states of the dopant atoms). The highest occupied state is indicated by the horizontal line. Upper panel gives the 3D plots of the integrated density of states. The integration was performed for the highest occupied states of the X-$(MgO)_{12}$ nanotubes within the shaded areas of the DOS plots.

### 3.2. CO adsorption on doped MgO nanotubes

As discussed above, the introduction of dopants into the structure of a MgO leads to the appearance of unpaired electrons, either at the dopant atoms (doping with B, C, and N) or surrounding O and Mg atoms (doping with Li and F). As suggested earlier [21], such sites can be considered as the confined radical species, the centers of altered chemical reactivity. This might open a number of possibilities for employing such doped oxide materials, for example, as adsorbents or heterogeneous catalysts. To explore this possibility, we investigated the adsorption properties of doped nanotubes using CO as a reactivity probe.

It is known that the MgO(001) surface is rather inert and it binds CO only weakly. The bonding is preferred at the Mg site with the adsorption energy of −0.16 eV [21]. This value agrees well with the experimental one, which is −0.14 eV [41]. For the considered non-doped hexagonal MgO nanotubes we find that CO prefers to bind at the edge Mg sites with the adsorption energies between −0.41 and −0.45 eV (PBE results, Table 2). Dispersion interactions have a small contribution to the adsorption energy (up to 10% of $E_{ads}$(CO)), similar to the CO adsorption on the MgO surface [42]. The calculated adsorption energies for the edge Mg sites depend weakly on the nanotube length (Table 2).

**Table 2**. Energetic and structural parameters for CO adsorption at the edge sites of the pristine $(MgO)_{3k}$ nanotubes ($k$ = 2, 3, 4 and 5). PBE results for the equilibrium bond distances are provided.

| | **$E_{ads}$(CO) / eV** | | | |
|---|---|---|---|---|
| **Nanotube size** | **PBE** | **PBE+D2** | **$d_{Mg-C}$ / Å** | **$d_{C-O}$ / Å** |
| **$(MgO)_6$** | −0.41 | −0.44 | 2.25 | 1.15 |
| **$(MgO)_9$** | −0.41 | −0.44 | 2.28 | 1.14 |
| **$(MgO)_{12}$** | −0.43 | −0.47 | 2.24 | 1.15 |
| **$(MgO)_{15}$** | −0.45 | −0.46 | 2.25 | 1.15 |

Our result is in agreement with the one reported in Ref. [26] where the CO adsorption energy at the edge Mg site of the $(MgO)_{12}$ nanotube was calculated to be −0.40 eV (PBE calculations). For the preferential binding sites we observe a small charge transfer to CO (< 0.1 e) and the elongation of the C–O bond length is by only ~0.01 Å, compared to the bond in an isolated CO molecule ($d_{C\text{-}O}$ = 1.14 Å). The calculated CO adsorption energies and equilibrium bond lengths for all the considered adsorption sites of the pristine nanotubes of different lengths are given in Table S7 (Supplementary Information). While the CO adsorption at the edge Mg site is stronger than on the MgO(001) surface, we find that the Mg sites at inner $(MgO)_3$ rings of the nanotubes bind CO weaker than pristine MgO(001) or do not bind it at all (Table S7, Supplementary Information). We further inspected the electronic structure of the CO@$(MgO)_{12}$ system (Fig. 7), and observed that the CO states overlapped in energy with the states at the bottom of the conduction band dominated by the states of the edge Mg atoms (see Fig. 3). We suggest that this is the reason for the weak interactions of CO with Mg sites away from the nanotube edges. In overall, the obtained results suggest that CO physisorbs on pristine $(MgO)_{3k}$ nanotubes.

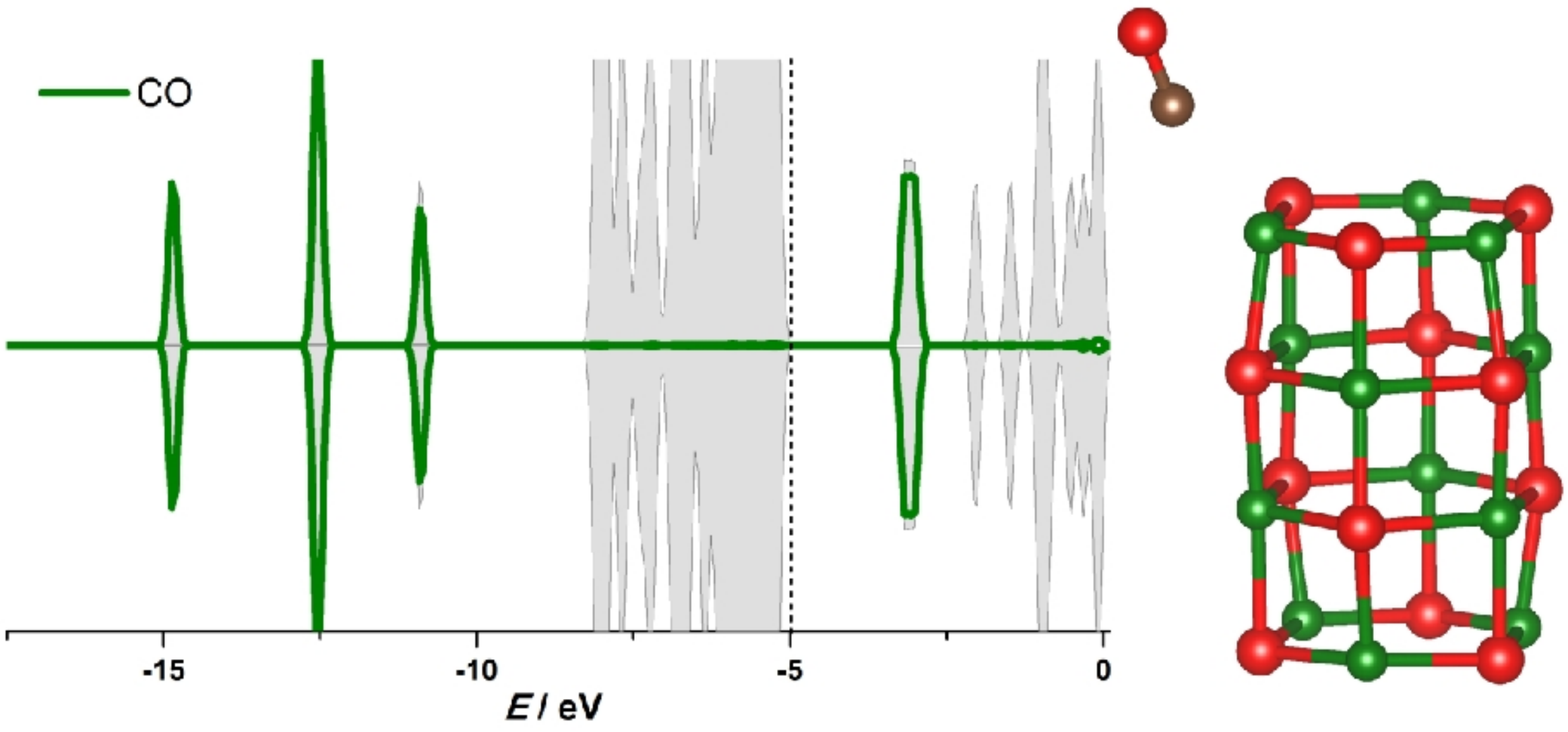

**Figure 7**. Projected densities of states of the CO molecule in the CO@$(MgO)_{12}$ system (CO adsorbed at the preferred edge Mg site). Total DOS (shaded) is also included. Vertical line indicates the highest occupied state. On the right the optimized structure of CO@$(MgO)_{12}$ is provided.

Further we investigate the CO adsorption on doped $(MgO)_{12}$ nanotubes in more detail. The results obtained for the preferential CO adsorption sites are given in Table 3. The full account of the CO adsorption energies and the corresponding bond lengths obtained for all the analyzed doped $(MgO)_{12}$ nanotubes is given in Tables S8-S12 (Supplementary Information). The notation for the adsorption sites is presented in Fig. 1.

**Table 3**. Energetic and structural parameters for CO adsorption on X-doped $(MgO)_{12}$ nanotubes.

| | | $E_{ads}$(CO) / eV | | | |
|---|---|---|---|---|---|
| **dopant** | **Ads. site** | **PBE** | **PBE+D2** | $d_{NT\text{-}C}$ / Å | $d_{C\text{-}O}$ / Å |
| **Li** | O01 | −1.80 | −1.79 | 1.26 | 1.22 |
| **B** | dopant | −3.74 | −3.79 | 1.43 | 1.19 |
| **C** | dopant | −5.19 | −5.29 | 1.30 | 1.20 |
| **N** | dopant | −3.06 | −3.12 | 1.23 | 1.18 |
| **F** | Mg01 | −0.78 | −0.82 | 2.20 | 1.17 |

In the cases of B-, C-, and N-doped nanotubes, where dopants bear unpaired electrons, the dopant sites are preferred for the CO adsorption. The interaction of CO with these sites is very strong. The adsorption energy reaches −5.19 eV for the case of C-doped $(MgO)_{12}$, indicating the formation of chemical bonds. For B- and C-doped $(MgO)_{12}$ the calculated CO adsorption energies are similar to those calculated for doped MgO(001) surfaces [21]. However, in the case of N-doped $(MgO)_{12}$ a strong interaction with the dopant site is observed. This is not the case for N-doped MgO(001) [21], for which a repulsive interaction between CO and the dopant is observed [21]. This lack of interaction is explained in terms of the orbital configuration, namely the 2*p*-orbitals of N lie in the surface plane, not overlapping with the CO orbitals. However, in the case of N-doped nanotubes the states of the dopant are available for bonding due to low coordination of the edge site, as can be seen from the integrated local density of states for this system (Fig. 6). We note that this situation occurs independent of the size of the N-doped MgO nanotube. Due to the filled electron shells of F and Li (formally being in the $F^-$ and $Li^+$ states) the preferred sites for the CO adsorption are now not the dopants themselves but their surrounding Mg and O atom, respectively. While the strong interaction between CO and the doped nanotubes can now be classified as chemisorption, we note that the CO adsorption at the other sites of doped MgO nanotubes remains rather weak (Tables S8-S12). However, for each considered dopant we find that the CO adsorption properties of the *entire* nanotube are altered when compared to those of the pristine $(MgO)_{12}$ nanotube. Namely, at all the considered adsorption sites CO binds stronger on doped nanotubes. By comparing the PBE and PBE+D2 results, a small contribution of dispersion interactions to $E_{ads}(CO)$ can be observed, being up to 0.15 eV (Table 3, Tables S8-S12). We can compare our results with those reported in Ref. [26] where the CO adsorption energies between −1.17 eV and −2.55 eV were calculated for Ni-, Pd- and Pt-doped $(MgO)_{12}$ nanotubes. The metal dopant replaced an Mg atom of the MgO nanotube and the CO adsorption was preferred at these dopant sites. We note that such a behavior is expected, being analogous to the cases of B-, C-, and N-doped nanotubes, as these metal atoms do not have fully filled electron shells.

Strong chemical interactions between the CO molecule and the doped nanotubes are also confirmed by the analysis of the electronic structure of the CO@X-$(MgO)_{12}$ systems (Fig. 8). In comparison to the CO@$(MgO)_{12}$ system (Fig. 7), the electronic structure of the doped nanotubes with the CO molecule is significantly altered. There is an overlap between the CO states and the nanotube atoms states. The adsorption is now accompanied by a pronounced charge redistribution, as in the case of doped MgO(001) surface [21]. Depending on the dopant type, the preferred adsorption site is either electron donor (B-, C-, and F-doped $(MgO)_{12}$) or electron acceptor (Li- and N-doped $(MgO)_{12}$). In the former case, a charge transfer from the dopant atoms to the CO molecule is observed, while in the latter, charge is transferred from the CO molecule to the nanotube. For B- and C-doped nanotubes the situation is analogous to the one observed for doped MgO(001) [21]. The electron donation properties of the F-doped MgO nanotube are due to the preferential binging of the CO molecule to the electron rich neighboring Mg site. The electron accepting properties of the Li-doped MgO nanotube are due to the CO bonding to the electron deficient O edge atom. We also find that the C–O bond is elongated in all the studied cases (Table 3), suggesting that the CO molecule is activated upon the adsorption. The charge transfer to/from the CO molecule makes it susceptible to the interactions with either electrophilic or nucleophilic species. Therefore, we suggest that the CO molecule can be activated towards oxidation or reduction by choosing the right type of dopant. Moreover, as the strength of the interaction of the reacting species with the catalyst is strongly connected to the catalyst performance [43], we suggest that the choice of the dopant could be used as a tool to tune the strength of the CO interaction with the MgO nanotube.

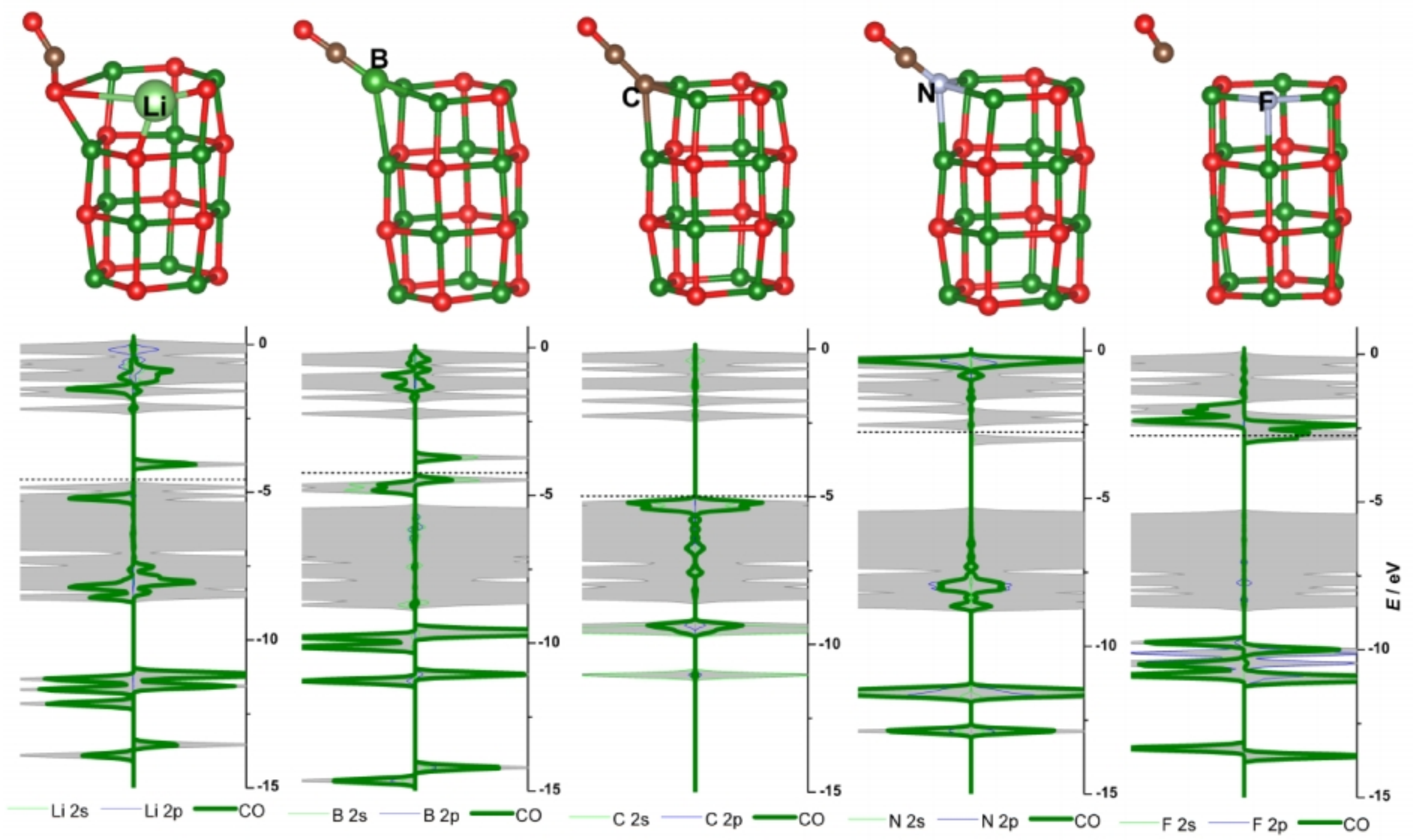

**Figure 8**. Projected densities of states of the CO molecule upon the adsorption at X-doped $(MgO)_{12}$ nanotubes (for the preferred adsorption sites). The total DOS (shaded) is included for each system. Horizontal dashed lines indicate the highest occupied states. The optimized structures of the CO@X-$(MgO)_{12}$ systems are included.

## 4. Conclusions

We have analyzed pristine and Li-, B-, C-, N-, and F-doped hexagonal MgO nanotubes. We find that the inclusion of dopants is preferred at the edges of MgO nanotubes, irrespective of their sizes. When doped, the MgO nanotubes are magnetic. The magnetization is due to the unpaired electron of the dopant atom (B-, C-, and N-doped MgO nanotubes) or its nearest O/Mg neighbors (Li- and F-doped MgO nanotubes). In addition, we find that the introduction of dopants alters the reactivity of the MgO nanotubes. While CO physisorbs on pristine MgO nanotubes ($E_{ads}$(CO) is between −0.41 eV and −0.45 eV), doped MgO nanotubes exhibit strong chemical interaction with CO, showing adsorption energies between −0.78 eV and −5.19 eV. Depending on the type of the dopant, the dopant site or their surrounding lattice atoms act as electron donors (B-, C-, and F-doped nanotubes) or electron acceptors (Li- and N-doped nanotubes). Hence, by choosing a dopant one can tune the strength of the interaction of CO with MgO nanotubes and also activate CO towards oxidation or reduction processes.

**Acknowledgement**

A.J., I.A.P. and M.P. acknowledge the support provided by the Serbian Ministry of Education, Science and Technological Development through the projects III45014 and 172040. N.V.S. acknowledges the support provided by Swedish Research Council through the project No. 2014-5993. We also acknowledge the support from Carl Tryggers Foundation for Scientific Research. The computations were performed on resources provided by the Swedish National Infrastructure for Computing (SNIC) at High Performance Computing Center North (HPC2N) at Umeå University. B3LYP calculations were performed on the PARADOX cluster at the Scientific Computing Laboratory of the Institute of Physics, Belgrade.

**References**

[1] C. Noguera, Physics and Chemistry at Oxide Surfaces, Cambridge University Press, Cambridge, 1996. doi:10.1017/CBO9780511524301.

[2] H.H. Kung, Transition metal oxides : surface chemistry and catalysis, Elsevier, 1989.

[3] M. Fernandez-Garcia, a Martinez-Arias, J.C. Hanson, J. a Rodriguez, Nanostructured Oxides in Chemistry:  Characterization and Properties, Chem. Rev. 104 (2004) 4063–4104. doi:doi:10.1021/cr030032f.

[4] K. Honkala, Tailoring oxide properties: An impact on adsorption characteristics of molecules and metals, Surf. Sci. Rep. 69 (2014) 366–388. doi:10.1016/j.surfrep.2014.09.002.

[5] M. Amft, N. V. Skorodumova, Catalytic activity of small MgO-supported Au clusters towards CO oxidation: A density functional study, Phys. Rev. B - Condens. Matter Mater. Phys. 81 (2010) 1–8. doi:10.1103/PhysRevB.81.195443.

[6] I. a. Pašti, N. V. Skorodumova, S. V. Mentus, Theoretical studies in catalysis and electrocatalysis: from fundamental knowledge to catalyst design, React. Kinet. Mech. Catal. (2014) 5–32. doi:10.1007/s11144-014-0808-x.

[7] B. Yoon, H. Häkkinen, U. Landman, A.S. Wörz, J.-M. Antonietti, S. Abbet, K.

Judai, U. Heiz, Charging effects on bonding and catalyzed oxidation of CO on Au8 clusters on MgO., Science. 307 (2005) 403–7. doi:10.1126/science.1104168.

[8] T. Lei, C. Ouyang, W. Tang, L.-F. Li, L.-S. Zhou, Preparation of MgO coatings on magnesium alloys for corrosion protection, Surf. Coatings Technol. 204 (2010) 3798–3803. doi:10.1016/j.surfcoat.2010.04.060.

[9] P. Poizot, S. Laruelle, S. Grugeon, L. Dupont, J.M. Tarascon, Nano-sized transition-metal oxides as negative-electrode materials for lithium-ion batteries., Nature. 407 (2000) 496–9. doi:10.1038/35035045.

[10] M. De Graef, M.E. McHenry, Crystal Structure Descriptions, Struct. Mater. (2012) 1–101.

[11] M.A. Shand, The chemistry and technology of magnesia, Wiley-Interscience, 2006.

[12] I. a. Pašti, M. Baljozović, N. V. Skorodumova, Adsorption of nonmetallic elements on defect-free MgO(001) surface – DFT study, Surf. Sci. 632 (2015) 39–49. doi:10.1016/j.susc.2014.09.012.

[13] I. a. Pašti, M.R. Baljozović, L.P. Granda-Marulanda, N. V. Skorodumova, Bimetallic dimers adsorbed on a defect-free MgO(001) surface: bonding, structure and reactivity, Phys. Chem. Chem. Phys. 17 (2015) 9666–9679. doi:10.1039/C4CP05723F.

[14] C.A. Scamehorn, N.M. Harrison, M.I. McCarthy, Water chemistry on surface defect sites: Chemidissociation versus physisorption on MgO(001), J. Chem. Phys. 101 (1994) 1547–1554. doi:10.1063/1.467777.

[15] S. Fernandez, A. Markovits, C. Minot, First Row Transition Metal Atom Adsorption On-Top of $F°_s$ Defects of a MgO(100) Surface, J. Phys. Chem. C. 112 (2008) 16491–16496. doi:10.1021/jp802166y.

[16] H.-J. Freund, G. Pacchioni, Oxide ultra-thin films on metals: new materials for the design of supported metal catalysts., Chem. Soc. Rev. 37 (2008) 2224–2242. doi:10.1039/b718768h.

[17] G. Pacchioni, N. Rösch, Supported nickel and copper clusters on MgO(100): A first-principles calculation on the metal/oxide interface, http://oasc12039.247realmedia.com/RealMedia/ads/click_lx.ads/www.aip.org/pt/adcenter/pdfcover_test/L-37/56140772/x01/AIP-PT/JCP_ArticleDL_110117/AIP-3075_JCP_Perspective_Generic_1640x440.jpg/434f71374e315a556e61414141774c75?x. (1998). doi:10.1063/1.471400.

[18] P.A. Žguns, M. Wessel, N. V. Skorodumova, Cu adatom charging on Mo supported ScN, MgO and NaF, RSC Adv. 5 (2015) 94436–94445. doi:10.1039/C5RA18565C.

[19] M. Grob, M. Pratzer, M. Morgenstern, M. Ležaić, Catalytic growth of N-doped MgO on Mo(001), Phys. Rev. B. 86 (2012) 75455. doi:10.1103/PhysRevB.86.075455.

[20] I.R. Shein, A.N. Enyashin, A.L. Ivanovskii, Magnetization of carbon-doped

MgO nanotubes, Phys. Rev. B - Condens. Matter Mater. Phys. 75 (2007) 1–5. doi:10.1103/PhysRevB.75.245404.

[21] I.A. Pašti, N. V. Skorodumova, Structural, electronic, magnetic and chemical properties of B-, C- and N-doped MgO(001) surfaces, Phys. Chem. Chem. Phys. 18 (2016) 426–435. doi:10.1039/C5CP05831G.

[22] I.A. Pašti, B. Johansson, N. V. Skorodumova, Tunable reactivity of supported single metal atoms by impurity engineering of the MgO(001) support, Phys. Chem. Chem. Phys. 20 (2018) 6337–6346. doi:10.1039/C7CP08370J.

[23] W.A. Saunders, Structural dissimilarities between small II-VI compound clusters: MgO and CaO, Phys. Rev. B. 37 (1988) 6583–6586. doi:10.1103/PhysRevB.37.6583.

[24] P.J. Ziemann, A.W. Castleman, Stabilities and structures of gas phase MgO clusters, J. Chem. Phys. 94 (1991) 718. doi:10.1063/1.460340.

[25] Y. Zhang, H.S. Chen, Y.H. Yin, Y. Song, Structures and bonding characters of $(MgO)_{3n}$ ( $n$ = 2–8) clusters, J. Phys. B At. Mol. Opt. Phys. 47 (2014) 25102. doi:10.1088/0953-4075/47/2/025102.

[26] M. Yang, Y. Zhang, S. Huang, H. Liu, P. Wang, H. Tian, Theoretical investigation of CO adsorption on TM-doped (MgO)12 (TM = Ni, Pd, Pt) nanotubes, Appl. Surf. Sci. 258 (2011) 1429–1436. doi:10.1016/J.APSUSC.2011.09.097.

[27] J.P. Perdew, K. Burke, M. Ernzerhof, D. of Physics, N.O.L. 70118 J. Quantum Theory Group Tulane University, Generalized Gradient Approximation Made Simple, Phys. Rev. Lett. 77 (1996) 3865–3868. doi:10.1103/PhysRevLett.77.3865.

[28] P. Giannozzi, S. Baroni, N. Bonini, M. Calandra, R. Car, C. Cavazzoni, D. Ceresoli, G.L. Chiarotti, M. Cococcioni, I. Dabo, A. Dal Corso, S. de Gironcoli, S. Fabris, G. Fratesi, R. Gebauer, U. Gerstmann, C. Gougoussis, A. Kokalj, M. Lazzeri, L. Martin-Samos, N. Marzari, F. Mauri, R. Mazzarello, S. Paolini, A. Pasquarello, L. Paulatto, C. Sbraccia, S. Scandolo, G. Sclauzero, A.P. Seitsonen, A. Smogunov, P. Umari, R.M. Wentzcovitch, QUANTUM ESPRESSO: a modular and open-source software project for quantum simulations of materials., J. Phys. Condens. Matter. 21 (2009) 395502. doi:10.1088/0953-8984/21/39/395502.

[29] S. Grimme, Semiempirical GGA-type density functional constructed with a long-range dispersion correction, J. Comput. Chem. 27 (2006) 1787–1799. doi:10.1002/jcc.20495.

[30] G.J. Martyna, M.E. Tuckerman, A reciprocal space based method for treating long range interactions in ab initio and force-field-based calculations in clusters, J. Chem. Phys. 110 (1999) 2810. doi:10.1063/1.477923.

[31] R.F.W. Bader, Atoms in molecules, 1990.

[32] G. Henkelman, A. Arnaldsson, H. Jónsson, A fast and robust algorithm for Bader decomposition of charge density, Comput. Mater. Sci. 36 (2006) 354–360. doi:10.1016/J.COMMATSCI.2005.04.010.

[33] M.J. et al Frisch, Gaussian 09, Revision D.01, Gaussian 09, Revis. D.01. (2009). doi:10.1159/000348293.

[34] C. Lee, W. Yang, R.G. Parr, Development of the Colle-Salvetti correlation-energy formula into a functional of the electron density, Phys. Rev. B. 37 (1988) 785–789. doi:10.1103/PhysRevB.37.785.

[35] A.D. Becke, Density-functional thermochemistry. III. The role of exact exchange, J. Chem. Phys. 98 (1993) 5648–5652. doi:10.1063/1.464913.

[36] T.H. Dunning, Gaussian basis sets for use in correlated molecular calculations. I. The atoms boron through neon and hydrogen, J. Chem. Phys. 90 (1989) 1007–1023. doi:10.1063/1.456153.

[37] R.A. Kendall, T.H. Dunning, R.J. Harrison, Electron affinities of the first-row atoms revisited. Systematic basis sets and wave functions, J. Chem. Phys. 96 (1992) 6796–6806. doi:10.1063/1.462569.

[38] S. Heo, E. Cho, H.-I. Lee, G.S. Park, H.J. Kang, T. Nagatomi, P. Choi, B.-D. Choi, Band gap and defect states of MgO thin films investigated using reflection electron energy loss spectroscopy, AIP Adv. 5 (2015) 77167. doi:10.1063/1.4927547.

[39] J.O. Nilsson, M. Leetmaa, B. Wang, P.A. Žguns, I. Pašti, A. Sandell, N. V Skorodumova, Modeling Kinetics of Water Adsorption on the Rutile TiO2 (110) Surface: Influence of Exchange-Correlation Functional, Phys. Status Solidi. (n.d.) 1700344--n/a. doi:10.1002/pssb.201700344.

[40] J.O. Nilsson, M. Leetmaa, B. Wang, P.A. Žguns, I. Pašti, A. Sandell, N. V. Skorodumova, Modeling Kinetics of Water Adsorption on the Rutile $TiO_2$ (110) Surface: Influence of Exchange-Correlation Functional (Phys. Status Solidi B 3/2018), Phys. Status Solidi. 255 (2018) 1870112. doi:10.1002/pssb.201870112.

[41] R. Wichtendahl, M. Rodriguez-Rodrigo, U. Härtel, H. Kuhlenbeck, H.-J. Freund, Thermodesorption of CO and NO from Vacuum-Cleaved NiO(100) and MgO(100), Phys. Status Solidi. 173 (1999) 93–100. doi:10.1002/(SICI)1521-396X(199905)173:1<93::AID-PSSA93>3.0.CO;2-4.

[42] P. Ugliengo, A. Damin, Are dispersive forces relevant for CO adsorption on the MgO(001) surface?, Chem. Phys. Lett. 366 (2002) 683–690. doi:10.1016/S0009-2614(02)01657-3.

[43] H. Knözinger, K. Kochloefl, Heterogeneous Catalysis and Solid Catalysts, in: Ullmann's Encycl. Ind. Chem., Wiley-VCH Verlag GmbH & Co. KGaA, Weinheim, Germany, 2003. doi:10.1002/14356007.a05_313.

[44] W. Humphrey, A. Dalke, K. Schulten, VMD: Visual molecular dynamics, J. Mol. Graph. 14 (1996) 33–38. doi:10.1016/0263-7855(96)00018-5.

[45] K. Momma, F. Izumi, IUCr, *VESTA* : a three-dimensional visualization system for electronic and structural analysis, J. Appl. Crystallogr. 41 (2008) 653–658. doi:10.1107/S0021889808012016.

# SUPPLEMENTARY INFORMATION

## S1. Structural parameters in pristine hexagonal $(MgO)_{3k}$ nanotubes

**Table S1**. Bond lengths for *k* = 2 with presented optimized (PBE) structure

| at1-at2 | PBE | PBE+D2 | B3LYP |
|---|---|---|---|
| 1-2 | 1.91 | 1.90 | 1.91 |
| 7-8 | 1.91 | 1.90 | 1.91 |
| 1-7 | 2.00 | 2.01 | 2.00 |

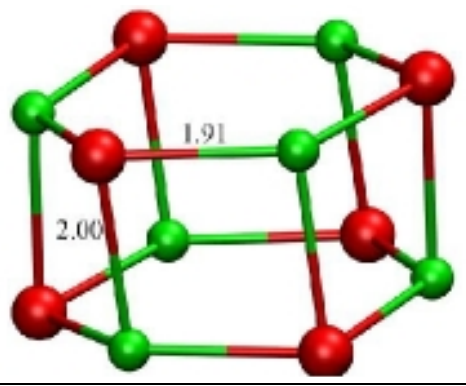

**Table S2**. Bond lengths for *k* = 3 with presented optimized (PBE) structure

| at1-at2 | PBE | PBE+D2 | B3LYP | at1-at2 | PBE | PBE+D2 | B3LYP |
|---|---|---|---|---|---|---|---|
| 1-2 | 1.90 | 1.89 | 1.91 | 1-7 | 1.94 | 1.95 | 1.95 |
| 7-8 | 2.06 | 2.05 | 2.06 | 2-8 | 2.01 | 2.01 | 2.01 |
| 13-14 | 1.90 | 1.89 | 1.91 | 3-9 | 1.94 | 1.95 | 1.95 |
| | | | | 4-10 | 2.01 | 2.01 | 2.01 |
| | | | | 5-11 | 1.94 | 1.95 | 2.95 |
| | | | | 6-12 | 2.01 | 2.01 | 2.01 |
| | | | | 7-13 | 1.94 | 1.95 | 1.95 |
| | | | | 8-14 | 2.01 | 2.01 | 2.01 |
| | | | | 9-15 | 1.94 | 1.95 | 1.95 |
| | | | | 10-16 | 2.01 | 2.01 | 2.01 |
| | | | | 11-17 | 1.94 | 1.95 | 1.95 |
| | | | | 12-16 | 2.01 | 2.01 | 2.01 |

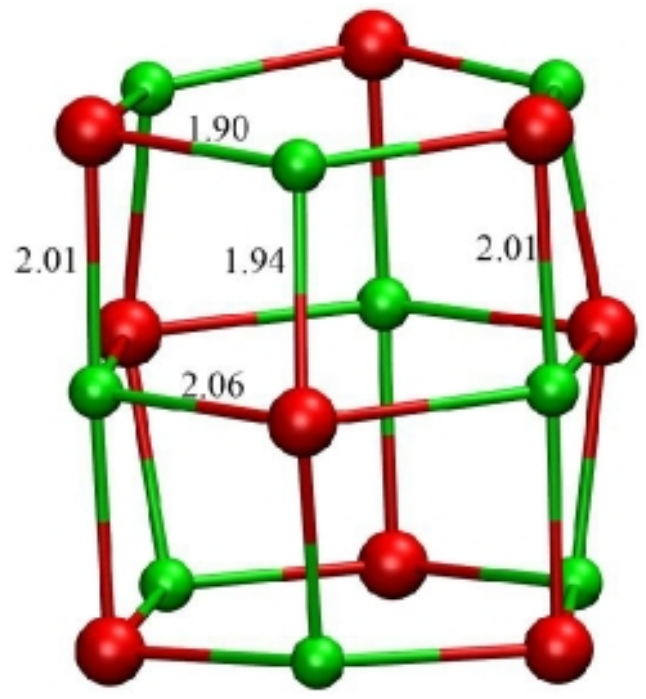

**Table S3**. Bond lengths for $k$ = 4 with presented optimized (PBE) structure

| at1-at2 | PBE | PBE+D2 | B3LYP | at1-at2 | PBE | PBE+D2 | B3LYP |
|---|---|---|---|---|---|---|---|
| 1-2 | 1.90 | 1.89 | 1.90 | 1-7 | 1.95 | 1.95 | 1.96 |
| 7-8 | 2.05 | 2.04 | 2.05 | 2-8 | 2.01 | 2.01 | 2.01 |
| 13-14 | 2.05 | 2.04 | 2.05 | 3-9 | 1.95 | 1.95 | 1.96 |
| 19-20 | 1.90 | 1.89 | 1.90 | 4-10 | 2.01 | 2.01 | 2.01 |
| | | | | 5-11 | 1.95 | 1.95 | 1.96 |
| | | | | 6-12 | 2.01 | 2.01 | 2.01 |
| | | | | 7-13 | 1.96 | 1.95 | 1.96 |
| | | | | 8-14 | 1.96 | 1.95 | 1.96 |
| | | | | 9-15 | 1.96 | 1.95 | 1.96 |
| | | | | 10-16 | 1.96 | 1.95 | 1.96 |
| | | | | 11-17 | 1.96 | 1.95 | 1.96 |
| | | | | 12-18 | 1.96 | 1.95 | 1.96 |
| | | | | 13-19 | 2.01 | 2.01 | 2.01 |
| | | | | 14-20 | 1.95 | 1.95 | 1.96 |
| | | | | 15-21 | 2.01 | 2.01 | 2.01 |
| | | | | 16-22 | 1.95 | 1.95 | 1.96 |
| | | | | 17-23 | 2.01 | 2.01 | 2.01 |
| | | | | 18-24 | 1.95 | 1.95 | 1.96 |

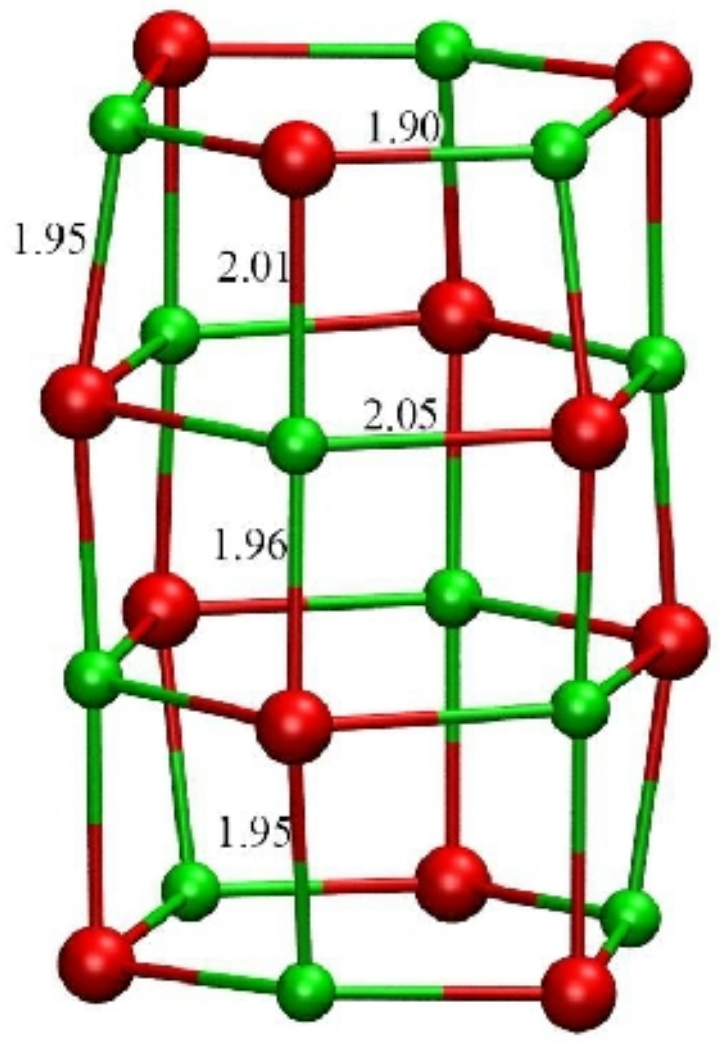

**Table S4**. Bond lengths for $k$ = 5 with presented optimized (PBE) structure

| at1-at2 | PBE | PBE+D2 | B3LYP | at1-at2 | PBE | PBE+D2 | B3LYP |
|---|---|---|---|---|---|---|---|
| 1-2 | 1.90 | 1.89 | 1.90 | 1-7 | 1.95 | 1.95 | 1.95 |
| 7-8 | 2.04 | 2.03 | 2.05 | 2-8 | 2.01 | 2.01 | 2.01 |
| 13-14 | 2.03 | 2.03 | 2.04 | 3-9 | 1.95 | 1.95 | 1.95 |
| 19-20 | 2.04 | 2.03 | 2.05 | 4-10 | 2.01 | 2.01 | 2.01 |
| 25-26 | 1.90 | 1.89 | 1.90 | 5-11 | 1.95 | 1.95 | 1.95 |
| | | | | 6-12 | 2.01 | 2.01 | 2.01 |
| | | | | 7-13 | 1.95 | 1.96 | 1.96 |
| | | | | 8-14 | 1.97 | 1.96 | 1.97 |
| | | | | 9-15 | 1.95 | 1.96 | 1.96 |
| | | | | 10-16 | 1.97 | 1.96 | 1.97 |
| | | | | 11-17 | 1.95 | 1.96 | 1.96 |
| | | | | 12-18 | 1.97 | 1.96 | 1.97 |
| | | | | 13-19 | 1.95 | 1.96 | 1.96 |
| | | | | 14-20 | 1.97 | 1.96 | 1.97 |
| | | | | 15-21 | 1.95 | 1.96 | 1.96 |
| | | | | 16-22 | 1.97 | 1.96 | 1.97 |
| | | | | 17-23 | 1.95 | 1.96 | 1.96 |
| | | | | 18-24 | 1.97 | 1.96 | 1.97 |
| | | | | 19-25 | 1.95 | 1.95 | 1.95 |
| | | | | 20-26 | 2.01 | 2.01 | 2.01 |
| | | | | 21-27 | 1.95 | 1.95 | 1.95 |
| | | | | 22-28 | 2.01 | 2.01 | 2.01 |
| | | | | 23-29 | 1.95 | 1.95 | 1.95 |
| | | | | 24-30 | 2.01 | 2.01 | 2.01 |

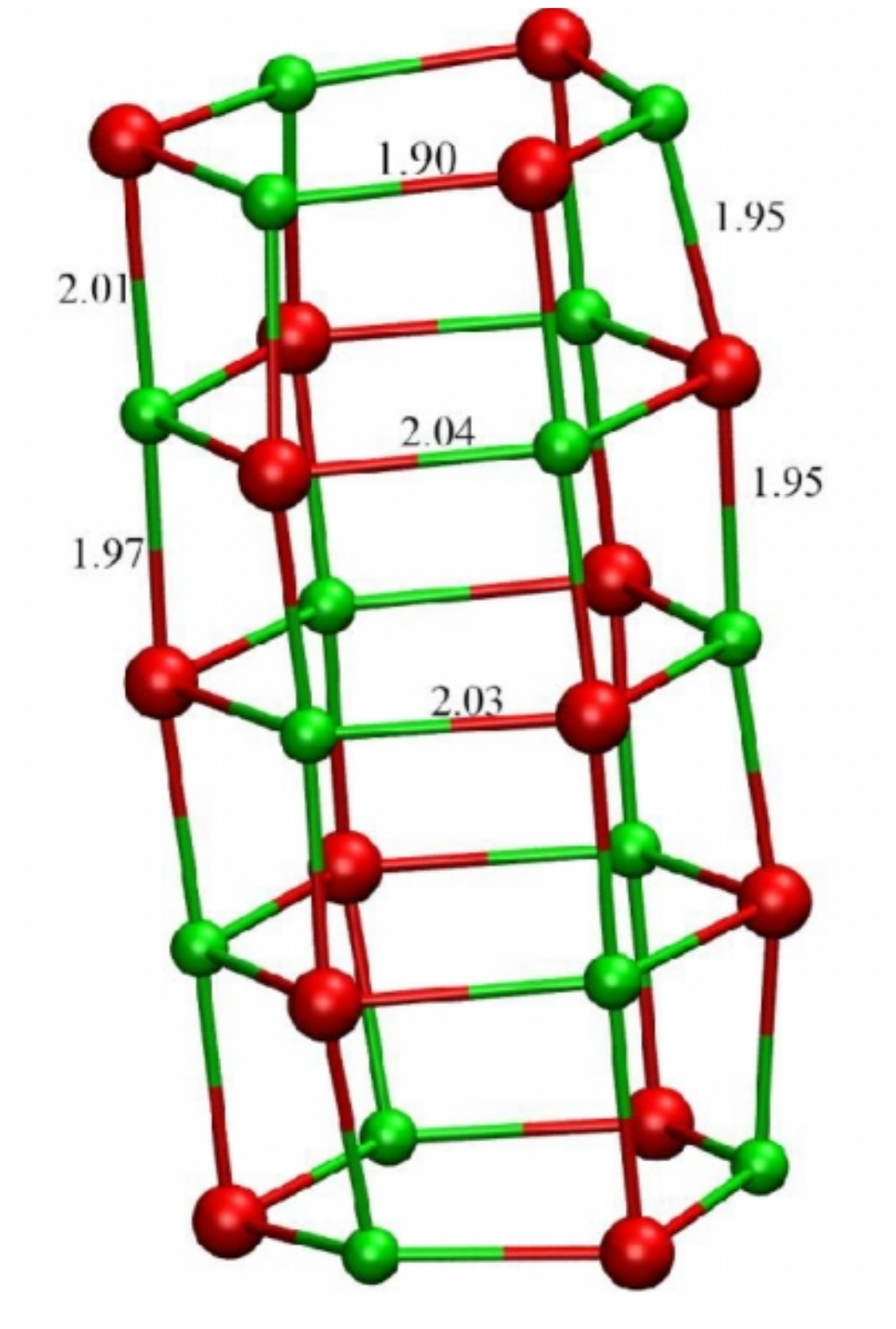

## S2. Edge saturation by H

We have calculated the H-saturated $(MgO)_{3k}$ clusters and investigated their electronic structure (Fig. S1). We also calculate the energetics of the saturation process with respect to $H_2$ ($E_{sat\text{-}H2}$) or $H_2O$ ($E_{sat\text{-}H2O}$) using the following equations:

$$E_{sat\text{-}H2} = (E_{H\text{-}MgO} - E_{MgO} - 3E_{H2})/3$$

$$E_{sat\text{-}H2O} = (E_{H\text{-}MgO} - E_{MgO} - 6E_{H2O} + 6E_{OH})/6$$

In the equations above $E_{H\text{-}MgO}$, $E_{MgO}$, $E_{H2}$, $E_{H2O}$ and $E_{OH}$ are the total energy of H-saturated $(MgO)_{3k}$, the total energy of bare $(MgO)_{3k}$, the total energy of the isolate $H_2$ molecule, the total energy of the isolated $H_2O$ molecule and the total energy of the isolated OH, respectively.

The results are summarized in Fig. S1.

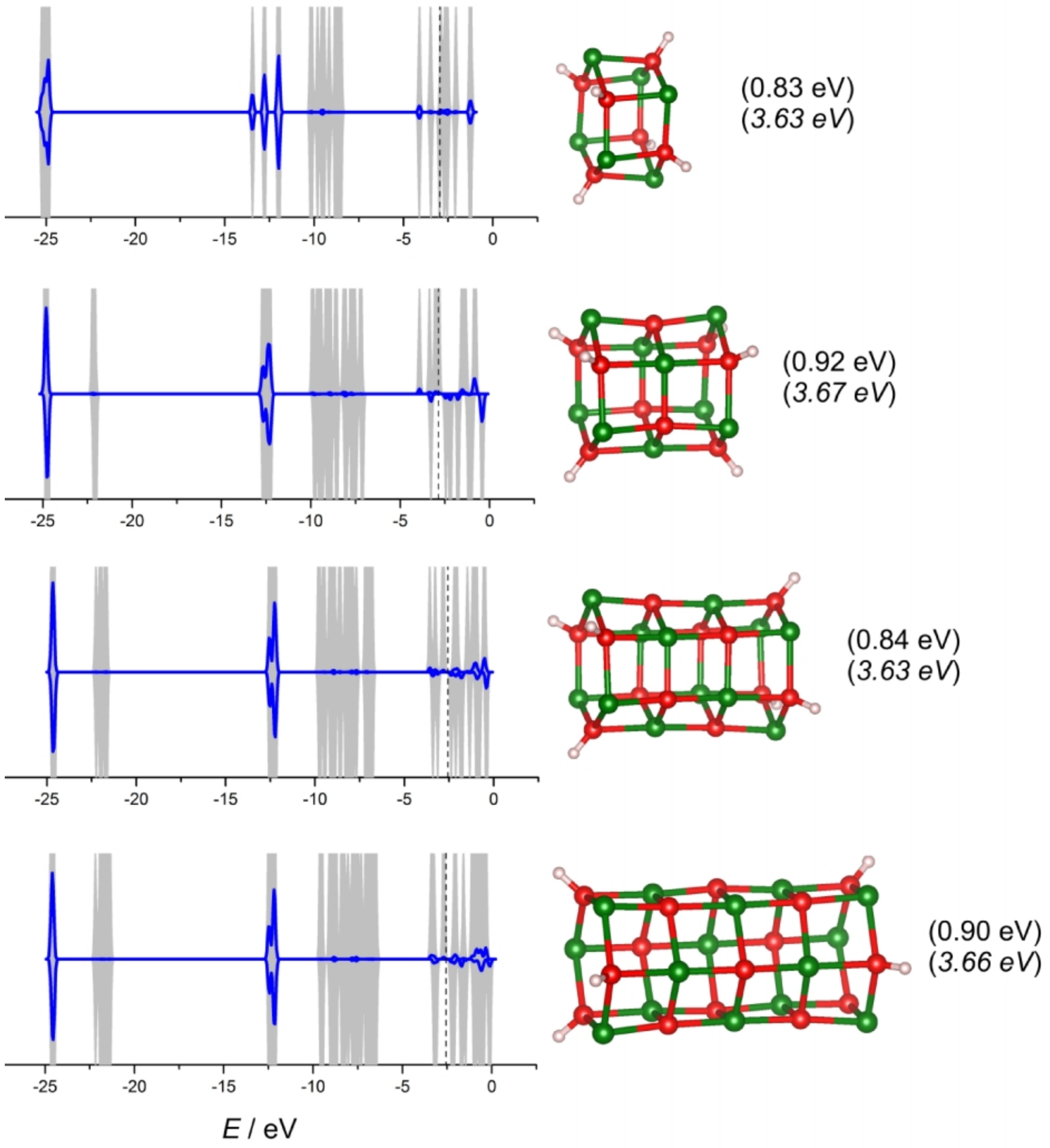

**Figure S1.** Projected densities of states of H-saturated $(MgO)_{3k}$ clusters (filled lines – total density of states, thick lines, H density of states) and the corresponding optimized structures. Numbers in parenthesis give the energy of the edge saturation with respect to the $H_2$ molecule ($E_{sat\text{-}H2}$) and $H_2O$ ($E_{sat\text{-}H2O}$, italic).

## S3. Vacancy formation energies

**Table S5**. Calculated vacancy formation energies for MgO nanotubes of different size and at different positions along a nanotube.

| | | $E_f^{vac}$ / eV oxygen vacancy | | | $E_f^{vac}$ / eV magnesium vacancy | | |
|---|---|---|---|---|---|---|---|
| | | **PBE** | **PBE+D2** | **B3LYP** | **PBE** | **PBE+D2** | **B3LYP** |
| **$(MgO)_6$** | **edge** | 7.94 | 7.94 | 8.36 | 5.22 | 5.44 | 7.34 |
| **$(MgO)_9$** | **edge** | 8.14 | 8.18 | 9.39 | 5.63 | 5.90 | 7.61 |
| | **2nd ring** | 8.35 | 8.44 | | 5.69 | 5.99 | |
| **$(MgO)_{12}$** | **edge** | 8.20 | 8.23 | 8.67 | 5.64 | 5.92 | 9.61 |
| | **2nd ring** | 8.61 | 8.72 | | 6.22 | 6.54 | |
| **$(MgO)_{15}$** | **edge** | 8.20 | 8.24 | 8.67 | 5.64 | 5.92 | 7.61 |
| | **2nd ring** | 8.66 | 8.77 | | 6.28 | 6.63 | |
| | **3rd ring** | 8.84 | 8.95 | | 6.68 | 7.02 | |

## S4. Adsorption energies of dopants at vacancy sites of MgO nanotubes

**Table S6**. Calculated binding energies at vacancy sites for considered dopants

| dopant | nanotube size | position | $E_b$(X) / eV PBE | PBE+D2 | B3LYP |
|---|---|---|---|---|---|
| Li | $(MgO)_6$ | edge | −4.49 | −4.68 | −5.38 |
| | $(MgO)_9$ | edge | −4.48 | −4.76 | −5.27 |
| | | $2^{nd}$ ring | −4.37 | −4.78 | |
| | $(MgO)_{12}$ | edge | −4.48 | −4.76 | −7.30 |
| | | $2^{nd}$ ring | −4.62 | −5.01 | |
| | $(MgO)_{15}$ | edge | −4.48 | −4.76 | −5.28 |
| | | $2^{nd}$ ring | −4.66 | −5.07 | |
| | | $3^{rd}$ ring | −4.82 | −5.18 | |
| B | $(MgO)_6$ | edge | −2.99 | −3.02 | −3.19 |
| | $(MgO)_9$ | edge | −3.08 | −3.13 | −4.12 |
| | | $2^{nd}$ ring | −2.89 | −3.01 | |
| | $(MgO)_{12}$ | edge | −3.11 | −3.15 | −3.36 |
| | | $2^{nd}$ ring | −3.02 | −3.14 | |
| | $(MgO)_{15}$ | edge | −3.11 | −3.16 | −3.37 |
| | | $2^{nd}$ ring | −3.02 | −3.14 | |
| | | $3^{rd}$ ring | −3.11 | −3.20 | |
| C | $(MgO)_6$ | edge | −4.39 | −4.40 | −4.82 |
| | $(MgO)_9$ | edge | −4.48 | −4.51 | −5.76 |
| | | $2^{nd}$ ring | −4.45 | −4.54 | |
| | $(MgO)_{12}$ | edge | −4.54 | −4.56 | −5.03 |
| | | $2^{nd}$ ring | −4.61 | −4.70 | |
| | $(MgO)_{15}$ | edge | −4.54 | −4.57 | −5.03 |
| | | $2^{nd}$ ring | −4.61 | −4.70 | |
| | | $3^{rd}$ ring | −4.73 | −4.81 | |
| N | $(MgO)_6$ | edge | −4.35 | −4.36 | −5.22 |
| | $(MgO)_9$ | edge | −4.51 | −4.55 | −6.22 |
| | | $2^{nd}$ ring | −4.50 | −4.61 | |
| | $(MgO)_{12}$ | edge | −4.55 | −4.58 | −5.48 |
| | | $2^{nd}$ ring | −4.70 | −4.80 | |
| | $(MgO)_{15}$ | edge | −4.56 | −4.59 | −5.47 |
| | | $2^{nd}$ ring | −4.73 | −4.84 | |
| | | $3^{rd}$ ring | −4.85 | −4.94 | |
| F | $(MgO)_6$ | edge | −6.44 | −6.45 | −7.18 |
| | $(MgO)_9$ | edge | −6.18 | −6.19 | −7.73 |
| | | $2^{nd}$ ring | −6.30 | −6.35 | |
| | $(MgO)_{12}$ | edge | −6.20 | −6.21 | −6.99 |
| | | $2^{nd}$ ring | −6.34 | −6.41 | |
| | $(MgO)_{15}$ | edge | −6.19 | −6.20 | −6.97 |
| | | $2^{nd}$ ring | −6.34 | −6.42 | |
| | | $3^{rd}$ ring | −6.39 | −6.44 | |

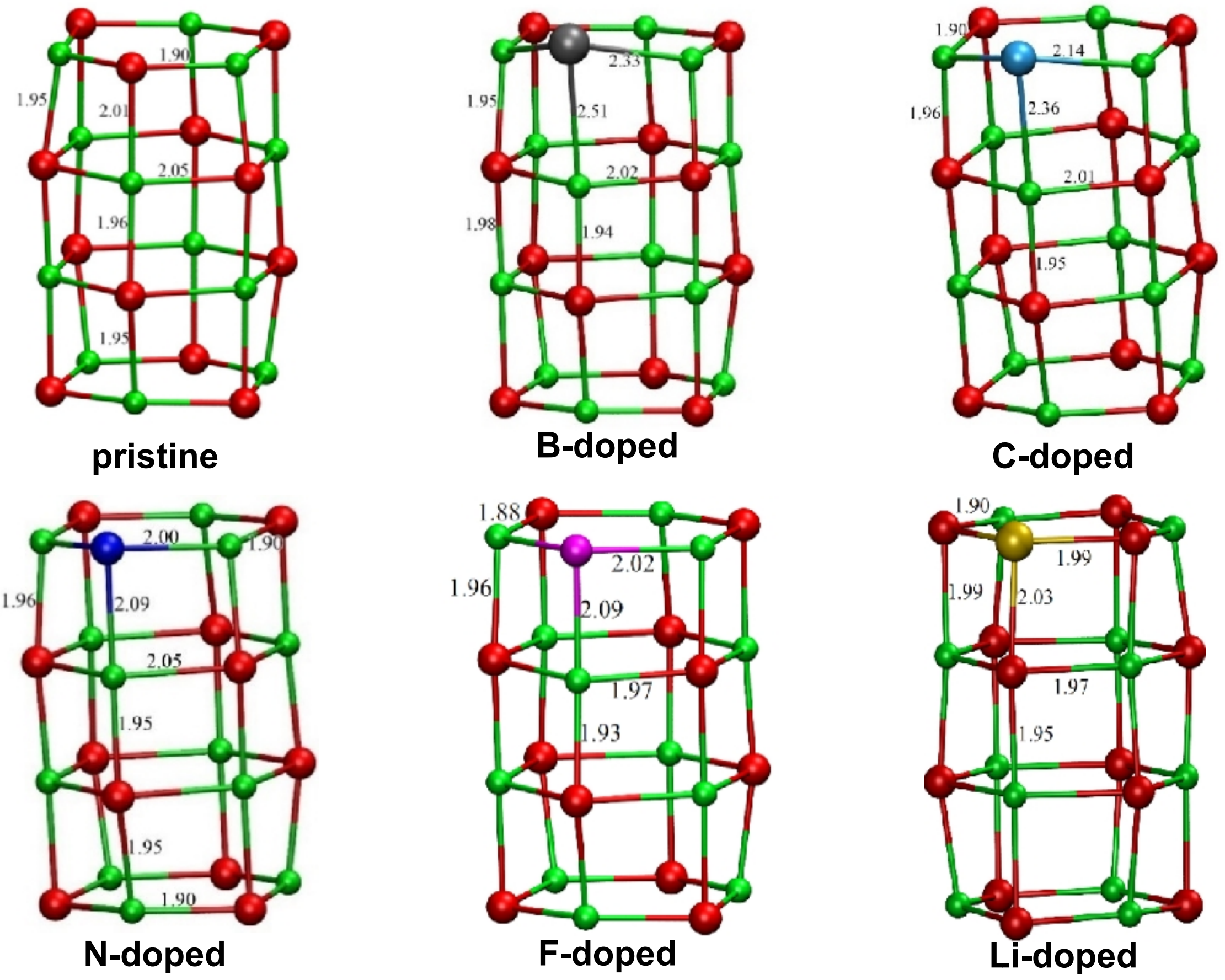

**Figure S2**. Optimized structures of pristine and doped $(MgO)_{12}$ nanotubes. Bond lengths are given in angstroms.

## S5. Charge density plots of doped MgO nanotubes

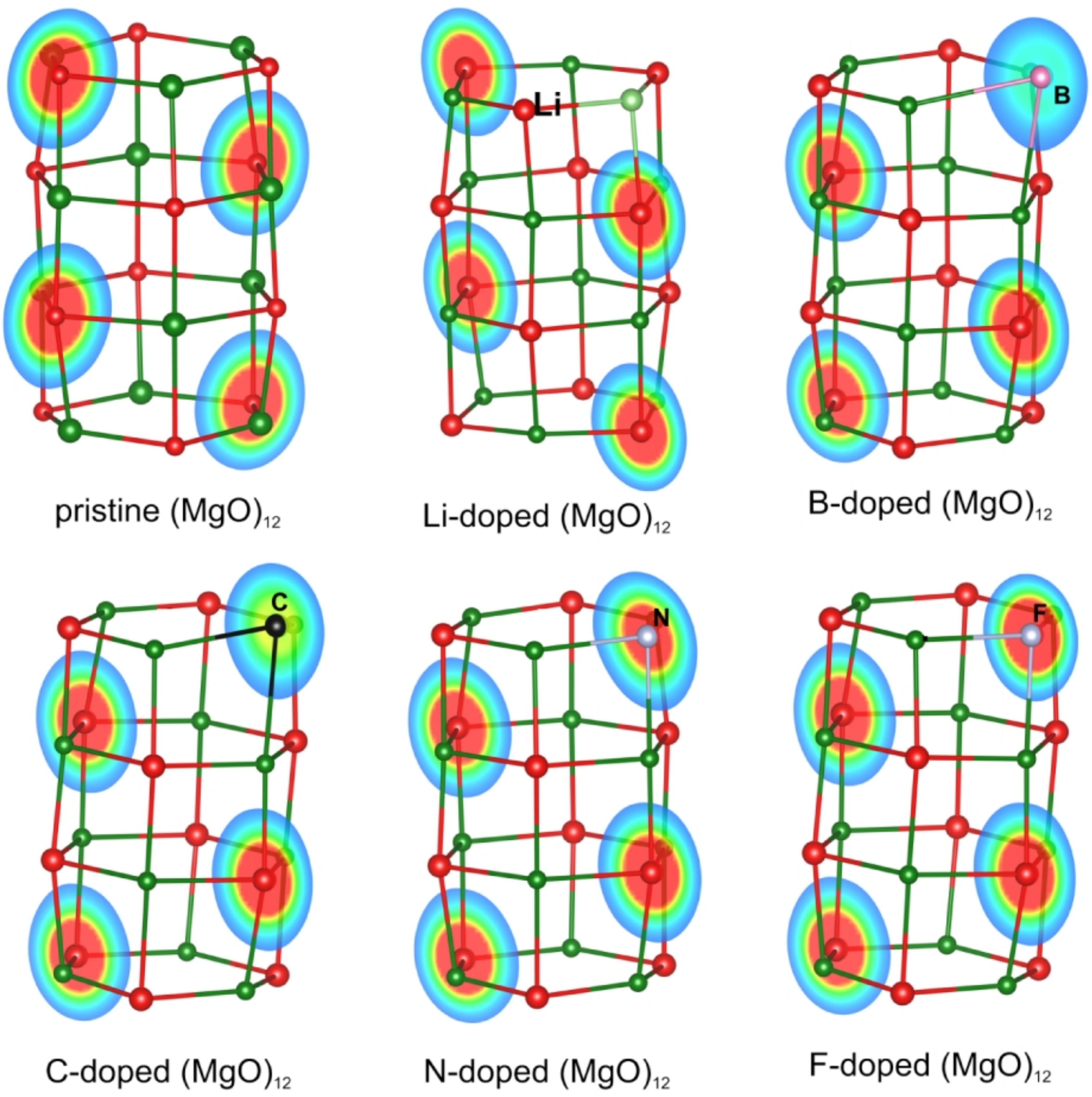

**Figure S3.** Charge density plots of pristine and doped (MgO)12 cluster with charge density projected on the symmetry planes passing through the dopant atoms.

## S6. Adsorption of CO on pristine hexagonal MgO nanotubes

**Table S7**. Energetic and structural parameters of CO adsorption on the pristine nanotubes. PBE results for the equilibrium bond distances are provided.

| | | $E_{ads}$(CO) / eV | | | |
|---|---|---|---|---|---|
| **nanotube size** | **Ads. site** | **PBE** | **PBE+D2** | **$d_{X\text{-}C}$ / Å** | **$d_{C\text{-}O}$ / Å** |
| **$(MgO)_6$** | **edge** | −0.41 | −0.44 | 2.25 | 1.15 |
| **$(MgO)_9$** | **edge** | −0.41 | −0.44 | 2.28 | 1.14 |
| | **2nd ring** | −0.01 | −0.09 | 2.30 | 1.15 |
| **$(MgO)_{12}$** | **edge** | −0.43 | −0.47 | 2.24 | 1.15 |
| | **2nd ring** | nb* | nb | | |
| **$(MgO)_{15}$** | **edge** | −0.45 | −0.46 | 2.25 | 1.15 |
| | **2nd ring** | nb | nb | | |
| | **3rd ring** | nb | nb | | |

*no binding, repulsive interaction

## S7. CO adsorption on doped $(MgO)_{12}$ nanotubes

**Table S8**. Adsorption of CO at different sites of Li-doped $(MgO)_{12}$ nanotube. The results for the preferential site are marked by bold letters.

| system/site | $E_{ads}$(CO) / eV PBE | $E_{ads}$(CO) / eV PBE+D2 | $d_{NT-C}$ / Å | $d_{C-O}$ / Å |
|---|---|---|---|---|
| CO@Li | −0.108 | −0.182 | 2.38 | 1.14 |
| CO@Mg01 | −0.430 | −0.468 | 2.33 | 1.14 |
| CO@Mg10 | −0.242 | −0.322 | 2.35 | 1.15 |
| CO@Mg11 | −0.108 | −0.193 | 2.27 | 1.15 |
| CO@Mg20 | −0.121 | −0.204 | 2.27 | 1.15 |
| CO@Mg21 | −0.289 | −0.371 | 2.35 | 1.14 |
| CO@Mg30 | −0.435 | −0.475 | 2.33 | 1.14 |
| CO@Mg31 | −0.436 | −0.473 | 2.34 | 1.14 |
| **CO@O01** | **−1.797** | **−1.793** | **1.26** | **1.22** |
| CO@O10 | −0.986 | −0.988 | 1.31 | 1.20 |

**Table S9**. Adsorption of CO at different sites of B-doped $(MgO)_{12}$ nanotube. The results for the preferential site are marked by bold letters.

| system/site | $E_{ads}$(CO) / eV PBE | $E_{ads}$(CO) / eV PBE+D2 | $d_{NT-C}$ / Å | $d_{C-O}$ / Å |
|---|---|---|---|---|
| **CO@B** | **−3.737** | **−3.788** | **1.43** | **1.19** |
| CO@Mg01 | −0.516 | ns* | 2.28 | 1.15 |
| CO@Mg02 | −0.484 | −0.580 | 2.32 | 1.14 |
| CO@Mg10 | ns | ns | / | / |
| CO@Mg11 | nb | −0.022 | 2.21 | 1.14 |
| CO@Mg20 | −0.273 | −0.356 | 2.26 | 1.15 |
| CO@Mg21 | −0.278 | −0.348 | 2.37 | 1.14 |
| CO@Mg30 | −0.475 | −0.506 | 2.34 | 1.14 |
| CO@Mg31 | −0.459 | −0.491 | 2.32 | 1.14 |

**Table S10**. Adsorption of CO at different sites of C-doped $(MgO)_{12}$ nanotube. The results for the preferential site are marked by bold letters.

| | $E_{ads}$(CO) / eV | | | |
|---|---|---|---|---|
| **system/site** | **PBE** | **PBE+D2** | $d_{NT-C}$ **/ Å** | $d_{C-O}$ **/ Å** |
| **CO@C** | **−5.194** | **−5.293** | **1.30** | **1.20** |
| **CO@Mg01** | −0.489 | −0.527 | 2.27 | 1.15 |
| **CO@Mg02** | −0.466 | −0.510 | 2.31 | 1.14 |
| **CO@Mg10** | ns | ns | / | / |
| **CO@Mg11** | −0.338 | −0.410 | 2.34 | 1.15 |
| **CO@Mg20** | −0.285 | −0.355 | 2.36 | 1.14 |
| **CO@Mg21** | −0.136 | −0.249 | 2.37 | 1.15 |
| **CO@Mg30** | −0.436 | −0.478 | 2.32 | 1.14 |
| **CO@Mg31** | −0.459 | −0.493 | 2.33 | 1.14 |

**Table S11**. Adsorption of CO at different sites of N-doped $(MgO)_{12}$ nanotube. The results for the preferential site are marked by bold letters.

| | $E_{ads}$(CO) / eV | | | |
|---|---|---|---|---|
| **system/site** | **PBE** | **PBE+D2** | $d_{NT-C}$ **/ Å** | $d_{C-O}$ **/ Å** |
| **CO@N** | **−3.059** | **−3.122** | **1.23** | **1.18** |
| **CO@Mg01** | −0.447 | −0.482 | 2.30 | 1.14 |
| **CO@Mg02** | −0.425 | −0.462 | 2.29 | 114 |
| **CO@Mg10** | ns | ns | / | / |
| **CO@Mg11** | −0.250 | −0.323 | 2.39 | 1.14 |
| **CO@Mg20** | −0.267 | −0.343 | 2.37 | 1.14 |
| **CO@Mg21** | −0.219 | −0.306 | 2.28 | 1.15 |
| **CO@Mg30** | −0.445 | −0.481 | 2.34 | 1.14 |
| **CO@Mg31** | −0.442 | −0.476 | 2.33 | 1.14 |

**Table S12**. Adsorption of CO at different sites of F-doped $(MgO)_{12}$ nanotube. The results for the preferential site are marked by bold letters.

| | $E_{ads}$(CO) / eV | | | |
|---|---|---|---|---|
| **system/site** | **PBE** | **PBE+D2** | $d_{NT-C}$ / Å | $d_{C-O}$ / Å |
| **CO@F** | ns | ns | / | / |
| **CO@Mg01** | **−0.781** | **−0.820** | **2.20** | **1.17** |
| **CO@Mg02** | −0.526 | −0.560 | 2.27 | 1.15 |
| **CO@Mg10** | −0.082 | −0.177 | 2.18 | 1.15 |
| **CO@Mg11** | −0.268 | −0.338 | 2.36 | 1.15 |
| **CO@Mg20** | −0.311 | −0.384 | 2.34 | 1.15 |
| **CO@Mg21** | −0.247 | −0.334 | 2.28 | 1.15 |
| **CO@Mg30** | −0.529 | −0.563 | 2.29 | 1.15 |
| **CO@Mg31** | −0.536 | −0.572 | 2.28 | 1.15 |